\documentclass[twocolumn,superscriptaddress,amsmath,amssymb,aps,prl]{revtex4-1}

\usepackage{graphicx}
\usepackage{dcolumn}
\usepackage{bm}

\usepackage{color}

\usepackage{graphicx}
\usepackage{comment}
\usepackage{color}
\usepackage[colorlinks,urlcolor=blue,citecolor=blue,linkcolor=blue]{hyperref}
\usepackage{cleveref}
\usepackage{physics}
\usepackage{mathtools}
\usepackage{mathrsfs}
\usepackage{booktabs}

\newcommand{\down}{\downarrow}
\newcommand{\up}{\uparrow}
\renewcommand{\k}{{\bf k}}

\newcommand{\q}{{\bf q}}
\newcommand{\0}{{\bf 0}}

\newcommand{\eq}{\epsilon_{\q}}

\newcommand{\ek}{\epsilon_{\k}}

\newcommand{\nn}{\nonumber}

\newcommand{\chd}{\hat c^\dag}
\newcommand{\ch}{\hat c}
\newcommand{\fhd}{\hat f^\dag}
\newcommand{\fh}{\hat f}
\newcommand{\dhd}{\hat d^\dag}
\renewcommand{\dh}{\hat d}

\newcommand{\jfl}[1]{{\color{black}#1}}
\newcommand{\fix}[1]{{\color{black}#1}}

\begin{document}

\title{Quasiparticle lifetime of the repulsive Fermi polaron}

\author{Haydn S. Adlong}
\affiliation{School of Physics and Astronomy, Monash University, Victoria 3800, Australia.}

\author{Weizhe Edward Liu}
\affiliation{School of Physics and Astronomy, Monash University, Victoria 3800, Australia.}
\affiliation{ARC Centre of Excellence in Future Low-Energy Electronics Technologies, Monash University, Victoria 3800, Australia}

\author{Francesco Scazza}
\affiliation{Istituto Nazionale di Ottica del Consiglio Nazionale delle Ricerche (CNR-INO) and European Laboratory for Nonlinear Spectroscopy (LENS), 50019 Sesto Fiorentino, Italy}

\author{Matteo Zaccanti}
\affiliation{Istituto Nazionale di Ottica del Consiglio Nazionale delle Ricerche (CNR-INO) and European Laboratory for Nonlinear Spectroscopy (LENS), 50019 Sesto Fiorentino, Italy}

\author{Nelson~\surname{Darkwah Oppong}}
\author{Simon F\"olling}
\affiliation{Ludwig-Maximilians-Universit{\"a}t, Schellingstra{\ss}e 4, 80799 M{\"u}nchen, Germany}
\affiliation{Max-Planck-Institut f{\"u}r Quantenoptik, Hans-Kopfermann-Stra{\ss}e 1, 85748 Garching, Germany}
\affiliation{Munich Center for Quantum Science and Technology (MCQST), Schellingstra{\ss}e 4, 80799 M{\"u}nchen, Germany}

\author{Meera M. Parish}
\affiliation{School of Physics and Astronomy, Monash University, Victoria 3800, Australia.}
\affiliation{ARC Centre of Excellence in Future Low-Energy Electronics Technologies, Monash University, Victoria 3800, Australia}

\author{Jesper Levinsen}
\affiliation{School of Physics and Astronomy, Monash University, Victoria 3800, Australia.}
\affiliation{ARC Centre of Excellence in Future Low-Energy Electronics Technologies, Monash University, Victoria 3800, Australia}

\date{\today}

\begin{abstract}
We investigate the metastable repulsive branch of a mobile impurity coupled to a degenerate Fermi gas via short-range interactions. 
We show that the quasiparticle lifetime of this repulsive Fermi polaron can be experimentally probed by driving Rabi oscillations between weakly and strongly interacting impurity states. Using a time-dependent variational approach, we find that we can accurately model the impurity Rabi oscillations that were recently measured for repulsive Fermi polarons in both two and three dimensions. Crucially, our theoretical description does not include relaxation processes to the lower-lying attractive branch.
Thus, the theory-experiment agreement demonstrates that the quasiparticle lifetime is \jfl{dominated} 
by many-body dephasing within the upper repulsive branch rather than by \jfl{relaxation from} 
the upper branch itself. Our findings shed light on recent experimental observations of persistent repulsive correlations, and have important consequences for the nature and stability of the strongly repulsive Fermi gas.
\end{abstract}

\maketitle

The concept of the quasiparticle is a powerful tool for describing interacting many-body quantum systems. Most notably, it forms the basis of Fermi liquid theory~\cite{Pines}, a highly successful phenomenological description of interacting Fermi systems ranging from liquid $^3$He to electrons in semiconductors. Here, the underlying particles are ``dressed'' by many-body excitations to form weakly interacting quasiparticles with modified properties such as a finite lifetime and an effective mass. However, during the past few decades, many materials have emerged that defy a conventional explanation within Fermi liquid theory~\cite{Varma2002,Norman2011}. Therefore, it is important to understand how quasiparticles can lose their coherence or break down.

Quantum impurities in quantum gases provide an ideal testbed in which to investigate quasiparticles since the impurity-medium interactions can be tuned to controllably create dressed impurity particles or \textit{polarons}~\cite{Massignan2014review}. To date, there have been a multitude of successful cold-atom experiments on impurities coupled to  
Fermi~\cite{Schirotzek2009,Nascimbene2009,Kohstall2012,Koschorreck2012,Zhang2012,Wenz2013,Cetina2015,Ong2015,Cetina2016,Scazza2017,Yan2019,Oppong2019,Ness2020} and Bose~\cite{Catani2012,Hu2016,Jorgensen2016,Camargo2018,Yan2019u} gases, which are termed Fermi and Bose polarons, respectively. In particular, Fermi-polaron experiments have observed the real-time formation of quasiparticles~\cite{Cetina2016} and the disappearance of quasiparticles in the spectral response with increasing temperature~\cite{Yan2019}. \jfl{The Fermi-polaron scenario has even been extended beyond cold atoms, having recently been realized in charge-tunable atomically thin semiconductors~\cite{Sidler2017}.} While the ground state of the Fermi polaron (corresponding to the attractive branch) is generally well understood~\cite{Chevy2006upd,Combescot2007,Prokofev2008,Combescot2008,Punk2009,Mathy2011,Schmidt2011,Trefzger2012,LevinsenResonant2DGases}, there has been much debate about the nature of the metastable repulsive branch~\cite{Cui2010,Pilati2010,Massignan2011,Schmidt2011,Goulko2016,Tajima2018,Mulkerin2019}, with experiments suggesting that it can be remarkably long-lived for a range of interactions~\cite{Kohstall2012,Scazza2017,Oppong2019}.
The stability of this branch is important for realizing Fermi gases with strong repulsive interactions~\cite{Massignan2014review,Pekker2011cbp,Sanner2012,Valtolina2017,Amico2018,Scazza2020}.

\begin{figure}
    \centering
    \includegraphics[width=0.9\columnwidth]{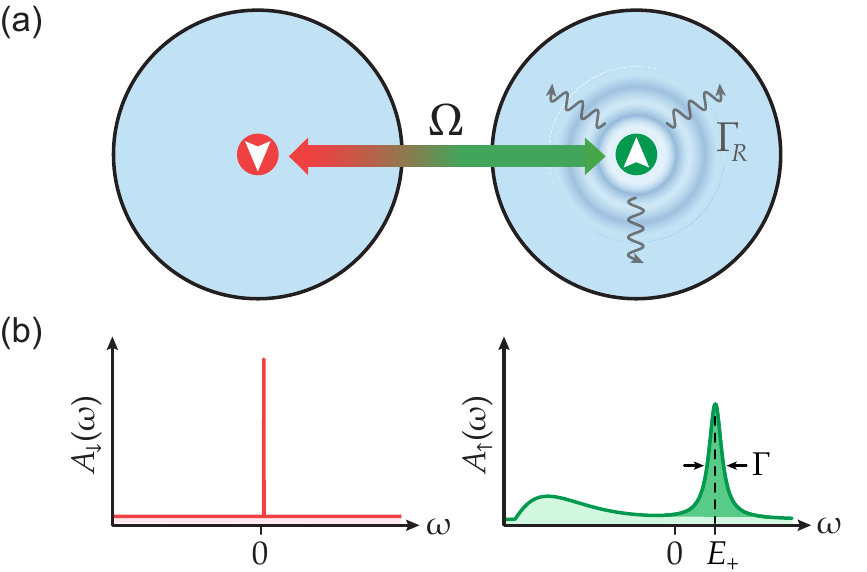}
    \caption{(a) Two pseudo-spin states (red and green) of an impurity embedded in a medium (blue) are coupled together and undergo Rabi oscillations with an effective frequency $\Omega$ and damping rate $\Gamma_R$. (b) The impurity spectral function of a nearly free impurity (left) is coupled to that of an impurity that strongly interacts with the Fermi gas (right). The repulsive polaron peak is centered at energy $E_+$ above the molecule-hole continuum and is characterized by the residue $Z$ (dark green area) and width $\Gamma$~\cite{Massignan2011}.}
    \label{fig:sketch}
    \vspace*{-20pt}
\end{figure}

In this Letter, we show that the lifetime of the repulsive branch itself is typically much longer than the \textit{quasiparticle lifetime}, \jfl{which corresponds to the time scale over which the repulsive polaron remains a coherent quasiparticle.} 
The character of the quasiparticle
can be probed by driving Rabi oscillations between different internal states of the impurity atom (see Fig.~\ref{fig:sketch}), where only one of the states strongly interacts with the surrounding Fermi gas.
Previous works have found that  
the Rabi frequency $\Omega$ provides a sensitive probe of the quasiparticle residue $Z$ (squared overlap with the non-interacting impurity state)~\cite{Kohstall2012}. Here we demonstrate that the damping rate of oscillations $\Gamma_R$ is directly linked to the width of the polaron peak in the spectral function, as depicted in Fig.~\ref{fig:sketch}(b), which corresponds to the inverse quasiparticle lifetime $\Gamma$.

Using a recently developed variational approach~\cite{Liu2019}, we model the Rabi oscillations 
for two different Fermi-polaron experimental setups: a three-dimensional (3D) $^6$Li gas with a broad Feshbach resonance~\cite{Scazza2017}, and a quasi-two-dimensional (2D) $^{173}$Yb gas~\cite{Oppong2019} with an orbital Feshbach resonance~\cite{Zhang2015}.  
We find that we can \jfl{capture the Rabi dynamics observed in both experiments, correctly reproducing both} 
$\Omega$ and $\Gamma_R$ 
even though our approximation neglects relaxation processes to the lower attractive branch at negative energies. 
We furthermore show that the repulsive polaron in the weak-coupling limit is essentially equivalent to the scenario of a discrete state coupled to a continuum, \jfl{which differs from the usual Fermi-liquid scenario.} %
Thus, we conclude that the quasiparticle lifetime of the repulsive Fermi polaron in two and three dimensions is primarily \jfl{limited by} many-body dephasing within the upper repulsive branch \jfl{while relaxation to the attractive branch is negligible}, in contrast to the prevailing wisdom~(see, e.g., Ref.~\cite{Massignan2014review} for a review).

\begin{figure*}
    \centering
    \includegraphics[width=1.85\columnwidth]{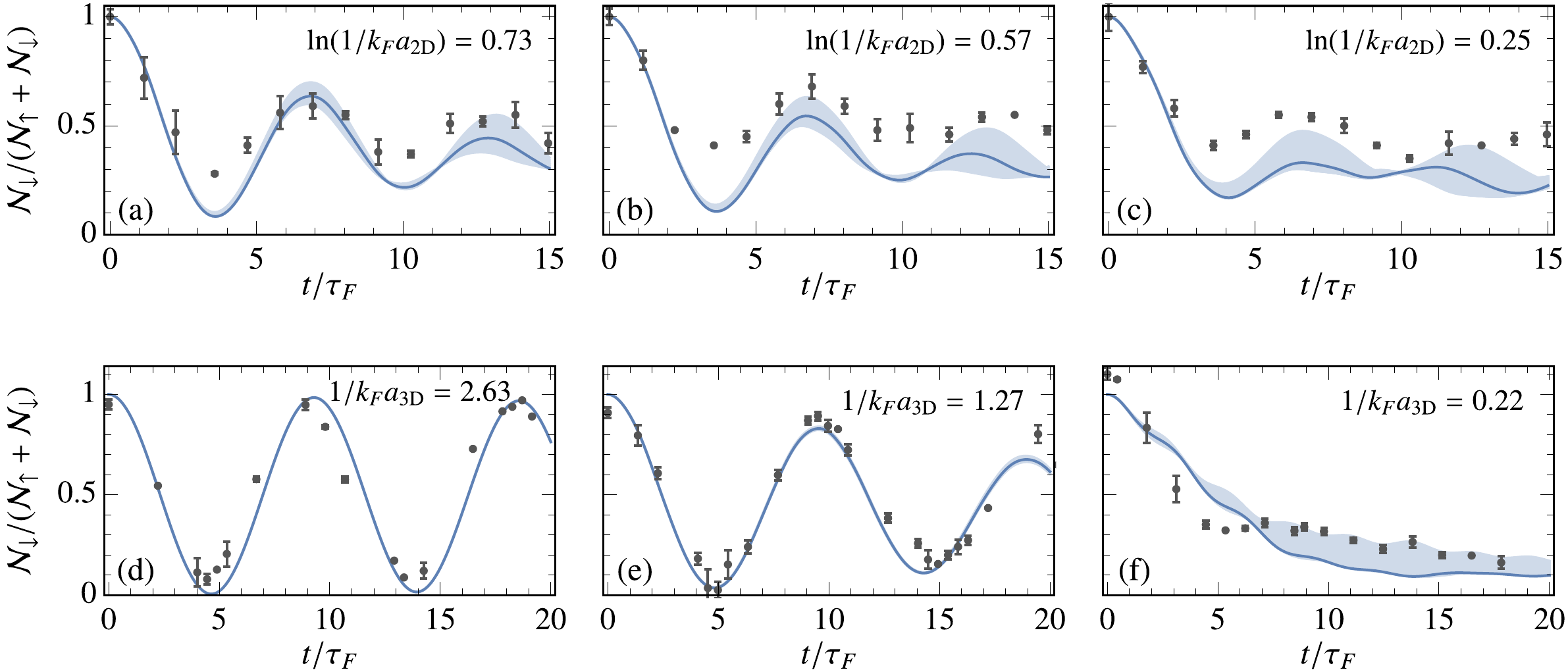}
    \caption{Rabi oscillations %
    calculated using the TBM (solid lines) and compared with the data (black dots) from the 2D ${}^{173}$Yb experiment~\cite{Oppong2019} (top row) and the 3D ${}^6$Li experiment~\cite{Scazza2017} (bottom row).  
    In the top row, $\Omega_0/E_F \simeq 0.95$ and $T/T_F \simeq 0.16$ while, from left to right, $\ln(1/k_F a_{\text{2D}}) = $ $0.73,\, 0.57,$ and $0.25$. In the bottom row, $\Omega_0/E_F \simeq 0.68$ and  $T/T_F \simeq 0.13$ while, from left to right, $1/(k_F a)=$ $2.63,\, 1.27,$ and $0.22$.
    The shaded regions correspond to the estimated uncertainty in the detuning around the repulsive polaron energy~\cite{supmat}.
    \jfl{In the top row, the data is calculated from the sole measurement of $\mathcal{N}_\downarrow(t)$, and the normalization of each data point to $\mathcal{N}_\downarrow(t=0)$~\cite{supmat}.}
    We define the Fermi time $\tau_F \equiv 1/E_F$.
    }
    \label{fig:Rabi}
\end{figure*}

\paragraph*{Model.---}

To model the impurity dynamics in the 3D $^6$Li experiment of
Ref.~\cite{Scazza2017} and the 2D $^{173}$Yb experiment of Ref.~\cite{Oppong2019}, we use a unified notation
where the dimensionality of momenta and sums are implicit. For clarity, even though we consider homonuclear systems, we introduce majority fermion creation operators $\fhd_\k$ and 
impurity creation operators $\chd_{\k,\sigma}$
with two different pseudo-spins $\sigma=\up,\down$ (Fig.~\ref{fig:sketch}). For a description of the precise relationship to atomic states in experiments, see the Supplemental Material \cite{supmat}. 

The Hamiltonian we consider consists of four terms:
\begin{align}
\hat H=\hat H_0+\hat H_\up+\hat H_\down+\hat H_\Omega.
\end{align}
The term $\hat H_0=\sum_\k(\ek-\mu) \hat f_\k^\dag\hat f_\k$ describes the medium
in the absence of the impurity. Here, $\k$ is
the particle momentum, $\ek=|\k|^2/2m\equiv k^2/2m$ is the kinetic energy, and $m$ is the
mass of both the fermions and the impurity (we work in units where $\hbar$
and the system volume or area are set to 1). We use a grand canonical formulation for the medium, with $\mu$ the corresponding chemical potential~\cite{Liu2020short,Liu2020long}.

The impurity spin-$\sigma$ terms 
\begin{align}
    \hat{H}_{\sigma} = & \sum_{\k} \left[\epsilon_{\k} \hat
                       c^\dag_{\k\sigma} \hat c_{\k\sigma} +
                       ( \ek/2 + \nu_{\sigma}) \hat d^\dag_{\k\sigma} \hat d_{\k\sigma}\right] \nn \\
                     &
+g_{\sigma}\sum_{\k, \q} \left( \hat d^\dag_{\q\sigma}\hat c_{\q - \k\sigma}  \hat f_{\k}
                       + \hat f^\dag_{\k} \hat c^\dag_{\q
                       -\k \sigma}\hat d_{\q\sigma} \right),   
\label{eq:Himp}
\end{align}
describe the interaction of the impurity and majority
fermions via the coupling into a closed channel with creation
operator $\dhd_{\k\sigma}$, where we have coupling constant \jfl{$g_\sigma$} and closed-channel
detuning $\nu_\sigma$. Renormalizing the model enables us to trade
the bare parameters of the model --- the detuning, the coupling constant, and an
ultraviolet momentum cutoff --- for the physical interaction
parameters
which parameterize the 2D and 3D impurity-majority fermion low-energy scattering amplitudes 
\begin{subequations}
\begin{align} 
     f_{\text{2D}\sigma}(k) &\simeq   \frac{4 \pi}{ - \ln (k^2 a^2_{\text{2D}\sigma}) + R^2_{\text{2D}\sigma}k^2+i \pi}, \label{fAmp2d}\\
     f_{\text{3D}\sigma}(k) &\simeq \frac{1}{-
                              a_{\text{3D}\sigma}^{-1} 
+ ik },\label{fAmp3d} 
\end{align}
\end{subequations}
namely the 2D and 3D scattering lengths, $a_{\text{2D}\sigma}$ and
$a_{\text{3D}\sigma}$, and a 2D range parameter $R_{\text{2D}\sigma}$~\cite{Levinsen2013,supmat}.
The presence of $R_{\text{2D}\sigma}$ in Eq.~\eqref{fAmp2d} allows us to model the strongly energy-dependent scattering close to the $^{173}$Yb orbital Feshbach resonance~\cite{Zhang2015,Hofer2015,Pagano2015} in a 2D geometry. Conversely, we can safely neglect effective range corrections for the broad resonance, 3D case of $^6$Li. In what follows, we take the impurity spin-$\up$ (spin-$\down$) state to be strongly (weakly) interacting with the medium~\cite{supmat}, as depicted in Fig.~\ref{fig:sketch}. To simplify notation, we therefore identify $a_{\rm 3D}\equiv a_{\rm 3D\up}$, $a_{\rm 2D}\equiv a_{\rm 2D\up}$, and $R_{\rm 2D}\equiv R_{\rm 2D\up}$.

The radio-frequency~\cite{Scazza2017} or optical~\cite{Oppong2019} fields that couple the impurities in states $\up$ and $\down$
are described within the rotating wave approximation:
\begin{align}
   \hat H_\Omega=\frac{\Omega_0}{2} \sum_{\k} \left( 
   \chd_{\k \down} \ch_{\k \up} + 
   \chd_{\k \up} \ch_{\k \down} \right)+\Delta\omega\,\hat n_{\down}.
\end{align}
Here, $\hat n_{\sigma}=\sum_\k\left(\hat c^\dag_{\k\sigma} \hat c_{\k\sigma}+\hat
d^\dag_{\k\sigma} \hat d_{\k\sigma}\right)$ is the spin-$\sigma$ impurity number operator, $\Omega_0$ is the (bare) Rabi coupling,  and $\Delta\omega$ is the 
detuning from the bare $\down$-$\up$ transition.

\paragraph{Perturbative analysis.---}

We can gain insight into the nature of the repulsive Fermi polaron 
by analyzing the quasiparticle peak in the spectrum 
at weak interactions and temperature $T=0$, \jfl{such that the polaron is at rest}. 
Focussing on the 3D case, in the limit $k_Fa_{3D}\ll1$ the polaron energy $E_+$ is given by the mean-field expression $E_+=\frac{4k_Fa_{\rm 3D}}{3\pi}E_F+O(a_{\rm 3D}^2)$~\cite{Bishop1973}, where $E_F=\frac{k_F^2}{2m}$ is the Fermi energy with $k_F$ the Fermi momentum. Thus, the quasiparticle state is pushed up into the continuum of scattering states that exists above zero energy in the case of attractive interactions. In particular, by performing a perturbative analysis~\cite{supmat}, we find that the broadening of the quasiparticle peak [Fig.~\ref{fig:sketch}(b)] is \jfl{dominated by} 
the coupling to this continuum, such that the leading order behavior is
\begin{align}
    \Gamma 
    \simeq\frac{8(k_Fa_{\rm 3D})^4}{9\pi^3}E_F\simeq 0.029\,(k_Fa_{\rm 3D})^4E_F.
    \label{eq:GammaPT}
\end{align}
This has two important consequences. First, at orders below $a_{\rm 3D}^4$, the quasiparticle behavior is indistinguishable from the case of truly repulsive interactions, \jfl{where the lifetime would be infinite~\cite{Bishop1973,Pilati2010}}. In this regime, the repulsive polaron is adiabatically connected to the non-interacting impurity.
Second, the \jfl{contribution to the quasiparticle width 
from relaxation to the attractive branch is negligible in this limit, since it} is dominated by three-body recombination~\cite{Scazza2017} \jfl{and} 
takes the form $\Gamma_3\simeq 0.025(k_Fa_{\rm 3D})^6E_F\, \jfl{\ll\Gamma}$~\cite{Petrov2003}.
This illustrates that --- within the perturbative regime --- the finite quasiparticle lifetime arises from many-body dephasing processes that are manifestly distinct from relaxation to negative-energy states.
Moreover, 
this cannot be viewed as momentum relaxation like in usual Fermi liquid theory~\cite{Pines}, since we are considering a \textit{zero-momentum} quasiparticle.

\paragraph{Rabi oscillations as a probe of quasiparticles.---}

We now argue that Rabi oscillations provide a sensitive probe of the repulsive polaron width (or quasiparticle lifetime). 
We focus on zero total momentum, since we are interested in decoherence effects beyond the standard momentum relaxation occurring in Fermi liquid theory~\cite{Pines}.
At times $t\geq0$, the impurity population in spin $\sigma$ is
\begin{align}
{\cal N}_\sigma(t)=\Trace[\hat\rho_0\hat c
(t)\hat n_\sigma \hat c
^\dag(t)],
\label{eq:Rabi}
\end{align}
where $\hat c
(t)$ is the impurity operator in the Heisenberg picture. Here, our initial state $\hat c
(t=0)=\hat c_{\0\down}$ is chosen such that ${\cal N}_\down(0)=1$ and ${\cal N}_\up(0)=0$. 
The trace is over all states of the medium in the absence of the impurity, and we use the thermal density matrix $\hat \rho_0=\exponential(-\beta \hat
H_0)/\Trace[\exponential (-\beta \hat H_0)]$ with $\beta\equiv1/T$ (we set the Boltzmann constant to 1). Under the assumption that \fix{the initial zero-momentum component dominates such that}
$\hat n_\down\simeq \hat c_{\0 \down}^\dag \hat c_{\0 \down}$,  
we find~\cite{supmat}
\begin{align} \label{Eq:RabiFromGreen}
   {\cal N}_\down(t) \simeq \int d \omega d \omega' \, \Tilde{A}_\downarrow (\omega) \Tilde{A}_\downarrow (\omega') e^{-i (\omega - \omega')t},
\end{align}
where $\Tilde{A}_\down(\omega)$ is the spin-$\down$ impurity spectral function in the presence of Rabi coupling. Taking the Rabi oscillations to be on resonance with the repulsive quasiparticle, i.e., $\Delta\omega=E_+$, we can furthermore approximate the spin-dependent impurity Green's functions in the absence of Rabi coupling as $G_\downarrow(\omega)\simeq1/(\omega-E_++i\eta)$ and $G_\uparrow(\omega) \simeq Z/(\omega-E_+ + i\Gamma)$, where $Z$ is the quasiparticle residue and $\eta$ is a convergence factor that implicitly carries the limit $\eta \to 0^+$. With these approximations and as long as $\Gamma\lesssim\sqrt{Z}\Omega_0$, we finally obtain~\cite{supmat}
\begin{align} \label{eq:MainApproxRabi}
    {\cal N}_\down(t) & \simeq e^{- \Gamma t} 
    \left[\frac12+\frac12 \cos \left(t \sqrt{\Omega _0^2
   Z-\Gamma^2}\right)\right].
\end{align}

Equation~\eqref{eq:MainApproxRabi} provides two valuable insights. First, the Rabi oscillation frequency $\Omega$ is related to the residue via $Z \simeq (\Omega^2 + \Gamma^2)/\Omega_0^2$, which provides a correction to the standard approximation of $Z \simeq (\Omega/\Omega_0)^2$~\cite{Kohstall2012}. Second, we see that the damping of Rabi oscillations is precisely the quasiparticle width $\Gamma$. This key result has been observed in experiment~\cite{Scazza2017,Oppong2019} but has previously lacked theoretical support.

\paragraph*{Variational approach.---}

We now turn to modelling the experimental Rabi oscillations. We apply the finite-temperature
variational approach developed in Ref.~\cite{Liu2019} in the context of 
Ramsey spectroscopy of impurities in a Fermi
sea~\cite{Cetina2016} (see also Ref.~\cite{Parish2016} for a related
zero-temperature approach). The idea in this 
truncated basis method (TBM) is to
introduce a time-dependent variational impurity operator \jfl{
$\hat{c}(t)=\hat{c}_\up(t)+\hat{c}_\down(t)$ that 
only
approximately satisfies the Heisenberg equation of motion.
This allows us to
introduce an error operator $\hat\epsilon(t)\equiv i\partial_t \hat{
{c}}(t)- \comm*{\hat{
{c}}(t)}{\hat H}$} and an associated error quantity
$\Delta(t)\equiv\Tr[\hat \rho_0 \hat\epsilon(t) \hat\epsilon^\dag(t)]$.
Our variational ansatz for the spin-$\sigma$ component of the impurity operator is inspired by the work of Chevy~\cite{Chevy2006upd} and corresponds to
\begin{align} \label{eq:varop}
  &\hat{
  {c}}_\sigma(t) \!=\! 
\alpha_{0}^\sigma(t)\hat
  c_{\0\sigma}\!+\!\sum_\k\alpha_{\k}^\sigma(t)\fhd_\k\dh_{\k\sigma}
\!+\!\sum_{\k,\q} \alpha_{\k\q}^\sigma(t) \fhd_{\q} \fh_{\k} \ch_{\q-\k \sigma}. 
\end{align}
The variational operator consists of three terms: 
the bare impurity, the impurity bound to a fermion in a closed-channel dimer, and the impurity with a particle-hole excitation. The time dependence is entirely contained within the variational
coefficients $\{\alpha\}$, allowing us to impose the minimization condition
$\pdv*{\Delta(t)}{\dot\alpha^{\sigma*}_j(t)}=0$. Since %
the Rabi coupling is suddenly turned on at $t=0$, we can use the stationary solutions obtained from a set of linear equations for
the expansion coefficients. This follows Ref.~\cite{Liu2019} with
straightforward modifications due to the two possible impurity spin states~\cite{supmat}. %

Following the application of an external driving field, we obtain the Rabi oscillations \jfl{within our variational approach via Eq.~\eqref{eq:Rabi}.} 
%
We show the resulting Rabi oscillations in Fig.~\ref{fig:Rabi} for a representative set of interaction strengths and temperatures $T/T_F$ in both two and three dimensions, where the Fermi temperature $T_F = E_F$. Here we set the detuning to match the theoretical repulsive polaron energy, with a small shift due to initial state interactions~\cite{supmat}. The shaded regions illustrate the range of possible results that can be obtained by varying the detuning within the width of the repulsive polaron quasiparticle peak~\cite{supmat}. This accounts for the Rabi oscillations being slightly off resonance in experiment due to the non-zero density of impurities, the density inhomogeneity within the trap, and other technical limitations.

Figure~\ref{fig:Rabi} demonstrates that our variational approach %
captures %
the Rabi oscillations between the bare impurity and the repulsive polaron in the 2D~\cite{Oppong2019} and 3D~\cite{Scazza2017} experiments.
\jfl{We note that there is a small positive offset in the 2D data~\cite{supmat}, which does not strongly affect the extracted Rabi parameters.}
Crucially, our variational ansatz \jfl{does not incorporate} 
any processes where the repulsive polaron decays into the attractive branch, %
because these involve additional particle-hole excitations~\cite{Massignan2011} which are neglected in Eq.~\eqref{eq:varop}.
Therefore, the consistency between our theoretical results and the experiments provides strong evidence that the decoherence in the Rabi oscillations --- and hence the \jfl{inverse} quasiparticle lifetime \jfl{$\Gamma$} --- is physically dominated by %
the coupling to the continuum at positive energies, rather than by relaxation to the attractive branch. Given the fundamental differences between the two experiments~\cite{Scazza2017,Oppong2019}, we expect this to be a generic feature of the mobile Fermi polaron \jfl{with short-range attractive interactions}.

\paragraph*{Quasiparticle properties.---}
We can further quantify the nature of the repulsive polaron by determining the frequency $\Omega$ and damping $\Gamma_R$ of the observed Rabi oscillations, which can be modelled approximately as~\cite{Scazza2017}:
\begin{align} \label{eq:TheoryFit}
    \mathcal{N}_{\downarrow} (t) \simeq b e^{-\Gamma_{\text{bg}}t} + (1-b) e^{- \Gamma_R t} \cos(\Omega t).
\end{align}
Here, $b$ is a dimensionless fitting parameter, while $\Gamma_{\text{bg}}$ can be regarded as a background decay rate of the spin-$\downarrow$ state. We see that Eq.~\eqref{eq:TheoryFit} reduces to our theoretical expression in Eq.~\eqref{eq:MainApproxRabi} if we set $b=1/2$ and $\Gamma_{\rm bg}  = \Gamma_R$. In practice, we find that $\Gamma_{\rm bg}  < \Gamma_R$ since there are scattering %
processes that can populate the spin-$\down$ state without contributing to the damping of oscillations, and these are neglected in our approximation~\eqref{Eq:RabiFromGreen} .

\begin{figure}
    \centering
    \includegraphics[width=\columnwidth]{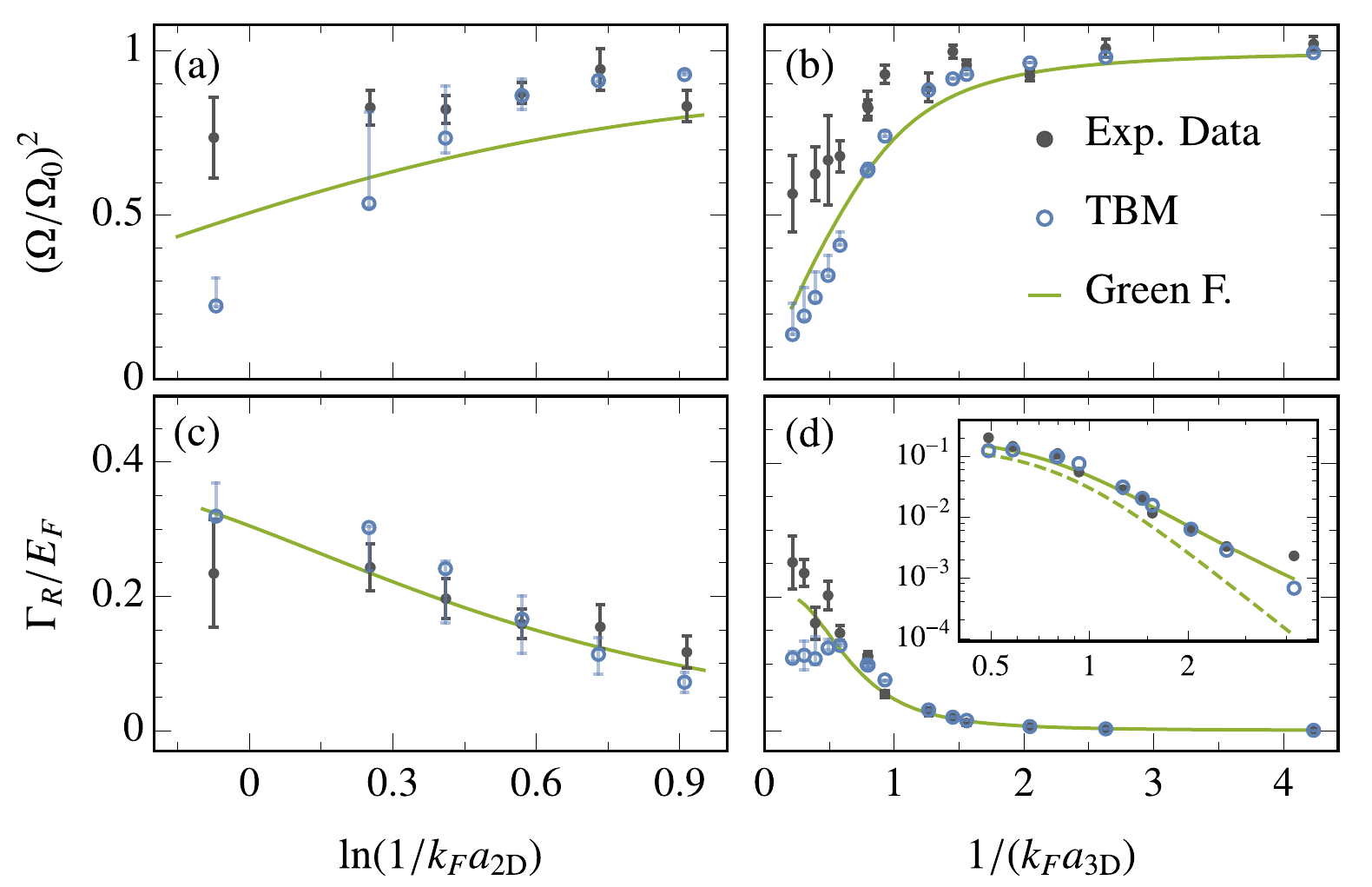}
    \caption{The extracted frequency $(\Omega/\Omega_0)^2$ and damping $\Gamma_R$ of the Rabi oscillations as determined from the TBM (blue circles) and in the 2D (a,c) and 3D (b,d) experiments (black dots). The TBM error bars are derived from the uncertainty in detuning~\cite{supmat}, \jfl{which tends to increase the oscillation frequency}. The extracted Rabi parameters are also compared with the \jfl{quasiparticle residue $Z\simeq (\Omega/\Omega_0)^2$ (top row) and quasiparticle inverse lifetime $\Gamma\simeq \Gamma_R$} (bottom row) obtained directly from the finite-temperature impurity Green's function (green solid line). 
    The TBM simulations are set to match the %
    experimental parameters~\cite{supmat}. %
    Inset: Data in (d) plotted on a log-log scale, together with the result of a zero-temperature Green's function calculation (green dashed line).}
    \label{fig:ResidueAndDamping}
\end{figure}

Using the fit provided in Eq.~\eqref{eq:TheoryFit}, we extract both $\Omega$ and $\Gamma_R$ from our simulated Rabi oscillations within the TBM and compare them with the experimental results, as depicted in Fig.~\ref{fig:ResidueAndDamping}.
We also 
\jfl{show} the repulsive polaron residue $Z$ and inverse quasiparticle lifetime $\Gamma$ \jfl{obtained} directly from the impurity Green's function $G_\down(\omega)$, where the impurity self energy is calculated using ladder diagrams~\cite{supmat}. Such an approach is equivalent to our spin-$\down$ variational ansatz %
in Eq.~\eqref{eq:varop} in the absence of Rabi coupling~\cite{Liu2019}, \jfl{and similarly it does not include the contribution to the quasiparticle lifetime due to relaxation from the repulsive branch to lower lying states.} 

Referring to Fig.~\ref{fig:ResidueAndDamping}, we see that all three methods are in good agreement with each other for weak to intermediate interaction strengths. 
This is particularly evident for $\Gamma_R$ in Fig.~\ref{fig:ResidueAndDamping}(d), where the agreement spans two orders of magnitude. Here we find that the finite temperature of the Fermi gas leads to deviations from the perturbative result in Eq.~\eqref{eq:GammaPT}, 
but %
the behavior is still markedly different from 
the momentum relaxation predicted by Fermi liquid theory~\cite{Cetina2015}. \jfl{The observed agreement suggests that temperature predominantly affects the many-body dephasing via thermal fluctuations of the medium rather than through finite impurity momenta.}
In Fig.~\ref{fig:ResidueAndDamping}(a), the slightly elevated values of $(\Omega/\Omega_0)^2$ compared to the expected residue $Z$ can be attributed to the strong Rabi driving ($\Omega_0/E_F \gtrsim 1$), 
such that the oscillation period approaches the formation time of the polaron.

At stronger repulsive interactions, there are more pronounced deviations as the repulsive polaron quasiparticle becomes less well defined \jfl{and the effects of a finite impurity density in experiments are expected to be more important~\cite{Scazza2017}}.
In particular, it becomes increasingly difficult to extract quasiparticle properties from the Rabi oscillations once the quasiparticle width $\Gamma$ is comparable to $\Omega$, which is consistent with our theoretical analysis in Eq.~\eqref{eq:MainApproxRabi}.
This accounts for the suppression of coherent oscillations in Fig.~\ref{fig:Rabi}(f) as well as the anomalously low $\Gamma_R$ obtained from the TBM near unitarity in Fig.~\ref{fig:ResidueAndDamping}(d). 
Our results suggest that one could better probe the repulsive polaron
quasiparticle lifetime at strong coupling by increasing the Rabi drive $\Omega_0$.

\paragraph*{Conclusions.---} We have investigated the nature of the repulsive Fermi polaron in two and three dimensions. We have shown that both the quasiparticle lifetime and the residue can be probed by driving Rabi oscillations between weakly and strongly interacting impurity spin states. By simulating the Rabi oscillations in two fundamentally different experiments and by performing a perturbative analysis of the weak-coupling limit,  we have demonstrated that  the quasiparticle lifetime is determined by many-body dephasing within the upper repulsive branch and is thus typically much shorter than the lifetime of the repulsive branch itself.
Our work \jfl{provides an important benchmark for many-body numerical approaches~\cite{Goulko2017}} and it opens up the prospect of exploring a long-lived repulsive Fermi gas with novel excitations beyond the Fermi liquid paradigm.

\acknowledgements 
We are grateful to J.~Cole, P.~Massignan, A.~Recati, and M.~Zonnios for useful discussions.  JL, WEL and MMP acknowledge support from the Australian Research Council Centre of Excellence in Future Low-Energy Electronics Technologies (CE170100039).  JL is also supported through the Australian Research Council Future Fellowship FT160100244.
NDO acknowledges funding from the International Max Planck Research School for Quantum Science and Technology. MZ was supported by the ERC through GA no.~637738 PoLiChroM. FS acknowledges funding from EU H2020 programme under the Marie Sk\l{}odowska-Curie GA no.~705269 and Fondazione Cassa di Risparmio di Firenze project \mbox{QuSim2D 2016.0770}.

\bibliography{polaron_bibliography}

\begin{thebibliography}{60}%
\makeatletter
\providecommand \@ifxundefined [1]{%
 \@ifx{#1\undefined}
}%
\providecommand \@ifnum [1]{%
 \ifnum #1\expandafter \@firstoftwo
 \else \expandafter \@secondoftwo
 \fi
}%
\providecommand \@ifx [1]{%
 \ifx #1\expandafter \@firstoftwo
 \else \expandafter \@secondoftwo
 \fi
}%
\providecommand \natexlab [1]{#1}%
\providecommand \enquote  [1]{``#1''}%
\providecommand \bibnamefont  [1]{#1}%
\providecommand \bibfnamefont [1]{#1}%
\providecommand \citenamefont [1]{#1}%
\providecommand \href@noop [0]{\@secondoftwo}%
\providecommand \href [0]{\begingroup \@sanitize@url \@href}%
\providecommand \@href[1]{\@@startlink{#1}\@@href}%
\providecommand \@@href[1]{\endgroup#1\@@endlink}%
\providecommand \@sanitize@url [0]{\catcode `\\12\catcode `\$12\catcode
  `\&12\catcode `\#12\catcode `\^12\catcode `\_12\catcode `\%12\relax}%
\providecommand \@@startlink[1]{}%
\providecommand \@@endlink[0]{}%
\providecommand \url  [0]{\begingroup\@sanitize@url \@url }%
\providecommand \@url [1]{\endgroup\@href {#1}{\urlprefix }}%
\providecommand \urlprefix  [0]{URL }%
\providecommand \Eprint [0]{\href }%
\providecommand \doibase [0]{http://dx.doi.org/}%
\providecommand \selectlanguage [0]{\@gobble}%
\providecommand \bibinfo  [0]{\@secondoftwo}%
\providecommand \bibfield  [0]{\@secondoftwo}%
\providecommand \translation [1]{[#1]}%
\providecommand \BibitemOpen [0]{}%
\providecommand \bibitemStop [0]{}%
\providecommand \bibitemNoStop [0]{.\EOS\space}%
\providecommand \EOS [0]{\spacefactor3000\relax}%
\providecommand \BibitemShut  [1]{\csname bibitem#1\endcsname}%
\let\auto@bib@innerbib\@empty
\bibitem [{\citenamefont {Pines}\ and\ \citenamefont
  {Nozi{\`e}res}(1966)}]{Pines}%
  \BibitemOpen
  \bibfield  {author} {\bibinfo {author} {\bibfnamefont {D.}~\bibnamefont
  {Pines}}\ and\ \bibinfo {author} {\bibfnamefont {P.}~\bibnamefont
  {Nozi{\`e}res}},\ }\href@noop {} {\emph {\bibinfo {title} {The Theory of
  Quantum Liquids}}}\ (\bibinfo  {publisher} {W. A. Benjamin},\ \bibinfo
  {address} {New York},\ \bibinfo {year} {1966})\BibitemShut {NoStop}%
\bibitem [{\citenamefont {Varma}\ \emph {et~al.}(2002)\citenamefont {Varma},
  \citenamefont {Nussinov},\ and\ \citenamefont {van Saarloos}}]{Varma2002}%
  \BibitemOpen
  \bibfield  {author} {\bibinfo {author} {\bibfnamefont {C.}~\bibnamefont
  {Varma}}, \bibinfo {author} {\bibfnamefont {Z.}~\bibnamefont {Nussinov}}, \
  and\ \bibinfo {author} {\bibfnamefont {W.}~\bibnamefont {van Saarloos}},\
  }\bibfield  {title} {\bibinfo {title} {\emph {Singular or non-Fermi
  liquids}},\ }\href {\doibase https://doi.org/10.1016/S0370-1573(01)00060-6}
  {\bibfield  {journal} {\bibinfo  {journal} {Physics Reports}\ }\textbf
  {\bibinfo {volume} {361}},\ \bibinfo {pages} {267 } (\bibinfo {year}
  {2002})}\BibitemShut {NoStop}%
\bibitem [{\citenamefont {Norman}(2011)}]{Norman2011}%
  \BibitemOpen
  \bibfield  {author} {\bibinfo {author} {\bibfnamefont {M.~R.}\ \bibnamefont
  {Norman}},\ }\bibfield  {title} {\bibinfo {title} {\emph {The Challenge of
  Unconventional Superconductivity}},\ }\href {\doibase
  10.1126/science.1200181} {\bibfield  {journal} {\bibinfo  {journal}
  {Science}\ }\textbf {\bibinfo {volume} {332}},\ \bibinfo {pages} {196}
  (\bibinfo {year} {2011})}\BibitemShut {NoStop}%
\bibitem [{\citenamefont {Massignan}\ \emph {et~al.}(2014)\citenamefont
  {Massignan}, \citenamefont {Zaccanti},\ and\ \citenamefont
  {Bruun}}]{Massignan2014review}%
  \BibitemOpen
  \bibfield  {author} {\bibinfo {author} {\bibfnamefont {P.}~\bibnamefont
  {Massignan}}, \bibinfo {author} {\bibfnamefont {M.}~\bibnamefont {Zaccanti}},
  \ and\ \bibinfo {author} {\bibfnamefont {G.~M.}\ \bibnamefont {Bruun}},\
  }\bibfield  {title} {\bibinfo {title} {\emph {Polarons, dressed molecules and
  itinerant ferromagnetism in ultracold Fermi gases}},\ }\href {\doibase
  10.1088/0034-4885/77/3/034401} {\bibfield  {journal} {\bibinfo  {journal}
  {Rep. Prog. Phys.}\ }\textbf {\bibinfo {volume} {77}},\ \bibinfo {pages}
  {034401} (\bibinfo {year} {2014})}\BibitemShut {NoStop}%
\bibitem [{\citenamefont {Schirotzek}\ \emph {et~al.}(2009)\citenamefont
  {Schirotzek}, \citenamefont {Wu}, \citenamefont {Sommer},\ and\ \citenamefont
  {Zwierlein}}]{Schirotzek2009}%
  \BibitemOpen
  \bibfield  {author} {\bibinfo {author} {\bibfnamefont {A.}~\bibnamefont
  {Schirotzek}}, \bibinfo {author} {\bibfnamefont {C.-H.}\ \bibnamefont {Wu}},
  \bibinfo {author} {\bibfnamefont {A.}~\bibnamefont {Sommer}}, \ and\ \bibinfo
  {author} {\bibfnamefont {M.~W.}\ \bibnamefont {Zwierlein}},\ }\bibfield
  {title} {\bibinfo {title} {\emph {{Observation of Fermi Polarons in a Tunable
  Fermi Liquid of Ultracold Atoms}}},\ }\href {\doibase
  10.1103/PhysRevLett.102.230402} {\bibfield  {journal} {\bibinfo  {journal}
  {Phys. Rev. Lett.}\ }\textbf {\bibinfo {volume} {102}},\ \bibinfo {pages}
  {230402} (\bibinfo {year} {2009})}\BibitemShut {NoStop}%
\bibitem [{\citenamefont {Nascimb{\`{e}}ne}\ \emph {et~al.}(2009)\citenamefont
  {Nascimb{\`{e}}ne}, \citenamefont {Navon}, \citenamefont {Jiang},
  \citenamefont {Tarruell}, \citenamefont {Teichmann}, \citenamefont
  {McKeever}, \citenamefont {Chevy},\ and\ \citenamefont
  {Salomon}}]{Nascimbene2009}%
  \BibitemOpen
  \bibfield  {author} {\bibinfo {author} {\bibfnamefont {S.}~\bibnamefont
  {Nascimb{\`{e}}ne}}, \bibinfo {author} {\bibfnamefont {N.}~\bibnamefont
  {Navon}}, \bibinfo {author} {\bibfnamefont {K.~J.}\ \bibnamefont {Jiang}},
  \bibinfo {author} {\bibfnamefont {L.}~\bibnamefont {Tarruell}}, \bibinfo
  {author} {\bibfnamefont {M.}~\bibnamefont {Teichmann}}, \bibinfo {author}
  {\bibfnamefont {J.}~\bibnamefont {McKeever}}, \bibinfo {author}
  {\bibfnamefont {F.}~\bibnamefont {Chevy}}, \ and\ \bibinfo {author}
  {\bibfnamefont {C.}~\bibnamefont {Salomon}},\ }\bibfield  {title} {\bibinfo
  {title} {\emph {{Collective Oscillations of an Imbalanced Fermi Gas: Axial
  Compression Modes and Polaron Effective Mass}}},\ }\href {\doibase
  10.1103/PhysRevLett.103.170402} {\bibfield  {journal} {\bibinfo  {journal}
  {Phys. Rev. Lett.}\ }\textbf {\bibinfo {volume} {103}},\ \bibinfo {pages}
  {170402} (\bibinfo {year} {2009})}\BibitemShut {NoStop}%
\bibitem [{\citenamefont {Kohstall}\ \emph {et~al.}(2012)\citenamefont
  {Kohstall}, \citenamefont {Zaccanti}, \citenamefont {Jag}, \citenamefont
  {Trenkwalder}, \citenamefont {Massignan}, \citenamefont {Bruun},
  \citenamefont {Schreck},\ and\ \citenamefont {Grimm}}]{Kohstall2012}%
  \BibitemOpen
  \bibfield  {author} {\bibinfo {author} {\bibfnamefont {C.}~\bibnamefont
  {Kohstall}}, \bibinfo {author} {\bibfnamefont {M.}~\bibnamefont {Zaccanti}},
  \bibinfo {author} {\bibfnamefont {M.}~\bibnamefont {Jag}}, \bibinfo {author}
  {\bibfnamefont {A.}~\bibnamefont {Trenkwalder}}, \bibinfo {author}
  {\bibfnamefont {P.}~\bibnamefont {Massignan}}, \bibinfo {author}
  {\bibfnamefont {G.}~\bibnamefont {Bruun}}, \bibinfo {author} {\bibfnamefont
  {F.}~\bibnamefont {Schreck}}, \ and\ \bibinfo {author} {\bibfnamefont
  {R.}~\bibnamefont {Grimm}},\ }\bibfield  {title} {\bibinfo {title} {\emph
  {{Metastability and coherence of repulsive polarons in a strongly interacting
  Fermi mixture}}},\ }\href {\doibase 10.1038/nature11065} {\bibfield
  {journal} {\bibinfo  {journal} {Nature}\ }\textbf {\bibinfo {volume} {485}},\
  \bibinfo {pages} {615} (\bibinfo {year} {2012})}\BibitemShut {NoStop}%
\bibitem [{\citenamefont {Koschorreck}\ \emph {et~al.}(2012)\citenamefont
  {Koschorreck}, \citenamefont {Pertot}, \citenamefont {Vogt}, \citenamefont
  {Fr{\"{o}}hlich}, \citenamefont {Feld},\ and\ \citenamefont
  {K{\"{o}}hl}}]{Koschorreck2012}%
  \BibitemOpen
  \bibfield  {author} {\bibinfo {author} {\bibfnamefont {M.}~\bibnamefont
  {Koschorreck}}, \bibinfo {author} {\bibfnamefont {D.}~\bibnamefont {Pertot}},
  \bibinfo {author} {\bibfnamefont {E.}~\bibnamefont {Vogt}}, \bibinfo {author}
  {\bibfnamefont {B.}~\bibnamefont {Fr{\"{o}}hlich}}, \bibinfo {author}
  {\bibfnamefont {M.}~\bibnamefont {Feld}}, \ and\ \bibinfo {author}
  {\bibfnamefont {M.}~\bibnamefont {K{\"{o}}hl}},\ }\bibfield  {title}
  {\bibinfo {title} {\emph {{Attractive and repulsive Fermi polarons in two
  dimensions}}},\ }\href {\doibase 10.1038/nature11151} {\bibfield  {journal}
  {\bibinfo  {journal} {Nature}\ }\textbf {\bibinfo {volume} {485}},\ \bibinfo
  {pages} {619} (\bibinfo {year} {2012})}\BibitemShut {NoStop}%
\bibitem [{\citenamefont {Zhang}\ \emph {et~al.}(2012)\citenamefont {Zhang},
  \citenamefont {Ong}, \citenamefont {Arakelyan},\ and\ \citenamefont
  {Thomas}}]{Zhang2012}%
  \BibitemOpen
  \bibfield  {author} {\bibinfo {author} {\bibfnamefont {Y.}~\bibnamefont
  {Zhang}}, \bibinfo {author} {\bibfnamefont {W.}~\bibnamefont {Ong}}, \bibinfo
  {author} {\bibfnamefont {I.}~\bibnamefont {Arakelyan}}, \ and\ \bibinfo
  {author} {\bibfnamefont {J.~E.}\ \bibnamefont {Thomas}},\ }\bibfield  {title}
  {\bibinfo {title} {\emph {{Polaron-to-Polaron Transitions in the
  Radio-Frequency Spectrum of a Quasi-Two-Dimensional Fermi Gas}}},\ }\href
  {\doibase 10.1103/PhysRevLett.108.235302} {\bibfield  {journal} {\bibinfo
  {journal} {Phys. Rev. Lett.}\ }\textbf {\bibinfo {volume} {108}},\ \bibinfo
  {pages} {235302} (\bibinfo {year} {2012})}\BibitemShut {NoStop}%
\bibitem [{\citenamefont {Wenz}\ \emph {et~al.}(2013)\citenamefont {Wenz},
  \citenamefont {Z{\"u}rn}, \citenamefont {Murmann}, \citenamefont {Brouzos},
  \citenamefont {Lompe},\ and\ \citenamefont {Jochim}}]{Wenz2013}%
  \BibitemOpen
  \bibfield  {author} {\bibinfo {author} {\bibfnamefont {A.~N.}\ \bibnamefont
  {Wenz}}, \bibinfo {author} {\bibfnamefont {G.}~\bibnamefont {Z{\"u}rn}},
  \bibinfo {author} {\bibfnamefont {S.}~\bibnamefont {Murmann}}, \bibinfo
  {author} {\bibfnamefont {I.}~\bibnamefont {Brouzos}}, \bibinfo {author}
  {\bibfnamefont {T.}~\bibnamefont {Lompe}}, \ and\ \bibinfo {author}
  {\bibfnamefont {S.}~\bibnamefont {Jochim}},\ }\bibfield  {title} {\bibinfo
  {title} {\emph {From Few to Many: Observing the Formation of a Fermi Sea One
  Atom at a Time}},\ }\href {\doibase 10.1126/science.1240516} {\bibfield
  {journal} {\bibinfo  {journal} {Science}\ }\textbf {\bibinfo {volume}
  {342}},\ \bibinfo {pages} {457} (\bibinfo {year} {2013})}\BibitemShut
  {NoStop}%
\bibitem [{\citenamefont {Cetina}\ \emph {et~al.}(2015)\citenamefont {Cetina},
  \citenamefont {Jag}, \citenamefont {Lous}, \citenamefont {Walraven},
  \citenamefont {Grimm}, \citenamefont {Christensen},\ and\ \citenamefont
  {Bruun}}]{Cetina2015}%
  \BibitemOpen
  \bibfield  {author} {\bibinfo {author} {\bibfnamefont {M.}~\bibnamefont
  {Cetina}}, \bibinfo {author} {\bibfnamefont {M.}~\bibnamefont {Jag}},
  \bibinfo {author} {\bibfnamefont {R.~S.}\ \bibnamefont {Lous}}, \bibinfo
  {author} {\bibfnamefont {J.~T.~M.}\ \bibnamefont {Walraven}}, \bibinfo
  {author} {\bibfnamefont {R.}~\bibnamefont {Grimm}}, \bibinfo {author}
  {\bibfnamefont {R.~S.}\ \bibnamefont {Christensen}}, \ and\ \bibinfo {author}
  {\bibfnamefont {G.~M.}\ \bibnamefont {Bruun}},\ }\bibfield  {title} {\bibinfo
  {title} {\emph {{Decoherence of Impurities in a Fermi Sea of Ultracold
  Atoms}}},\ }\href {\doibase 10.1103/PhysRevLett.115.135302} {\bibfield
  {journal} {\bibinfo  {journal} {Phys. Rev. Lett.}\ }\textbf {\bibinfo
  {volume} {115}},\ \bibinfo {pages} {135302} (\bibinfo {year}
  {2015})}\BibitemShut {NoStop}%
\bibitem [{\citenamefont {Ong}\ \emph {et~al.}(2015)\citenamefont {Ong},
  \citenamefont {Cheng}, \citenamefont {Arakelyan},\ and\ \citenamefont
  {Thomas}}]{Ong2015}%
  \BibitemOpen
  \bibfield  {author} {\bibinfo {author} {\bibfnamefont {W.}~\bibnamefont
  {Ong}}, \bibinfo {author} {\bibfnamefont {C.}~\bibnamefont {Cheng}}, \bibinfo
  {author} {\bibfnamefont {I.}~\bibnamefont {Arakelyan}}, \ and\ \bibinfo
  {author} {\bibfnamefont {J.~E.}\ \bibnamefont {Thomas}},\ }\bibfield  {title}
  {\bibinfo {title} {\emph {Spin-Imbalanced Quasi-Two-Dimensional Fermi
  Gases}},\ }\href {\doibase 10.1103/PhysRevLett.114.110403} {\bibfield
  {journal} {\bibinfo  {journal} {Phys. Rev. Lett.}\ }\textbf {\bibinfo
  {volume} {114}},\ \bibinfo {pages} {110403} (\bibinfo {year}
  {2015})}\BibitemShut {NoStop}%
\bibitem [{\citenamefont {Cetina}\ \emph {et~al.}(2016)\citenamefont {Cetina},
  \citenamefont {Jag}, \citenamefont {Lous}, \citenamefont {Fritsche},
  \citenamefont {Walraven}, \citenamefont {Grimm}, \citenamefont {Levinsen},
  \citenamefont {Parish}, \citenamefont {Schmidt}, \citenamefont {Knap},\ and\
  \citenamefont {Demler}}]{Cetina2016}%
  \BibitemOpen
  \bibfield  {author} {\bibinfo {author} {\bibfnamefont {M.}~\bibnamefont
  {Cetina}}, \bibinfo {author} {\bibfnamefont {M.}~\bibnamefont {Jag}},
  \bibinfo {author} {\bibfnamefont {R.~S.}\ \bibnamefont {Lous}}, \bibinfo
  {author} {\bibfnamefont {I.}~\bibnamefont {Fritsche}}, \bibinfo {author}
  {\bibfnamefont {J.~T.~M.}\ \bibnamefont {Walraven}}, \bibinfo {author}
  {\bibfnamefont {R.}~\bibnamefont {Grimm}}, \bibinfo {author} {\bibfnamefont
  {J.}~\bibnamefont {Levinsen}}, \bibinfo {author} {\bibfnamefont {M.~M.}\
  \bibnamefont {Parish}}, \bibinfo {author} {\bibfnamefont {R.}~\bibnamefont
  {Schmidt}}, \bibinfo {author} {\bibfnamefont {M.}~\bibnamefont {Knap}}, \
  and\ \bibinfo {author} {\bibfnamefont {E.}~\bibnamefont {Demler}},\
  }\bibfield  {title} {\bibinfo {title} {\emph {{Ultrafast many-body
  interferometry of impurities coupled to a Fermi sea}}},\ }\href {\doibase
  10.1126/science.aaf5134} {\bibfield  {journal} {\bibinfo  {journal}
  {Science}\ }\textbf {\bibinfo {volume} {354}},\ \bibinfo {pages} {96}
  (\bibinfo {year} {2016})}\BibitemShut {NoStop}%
\bibitem [{\citenamefont {Scazza}\ \emph {et~al.}(2017)\citenamefont {Scazza},
  \citenamefont {Valtolina}, \citenamefont {Massignan}, \citenamefont {Recati},
  \citenamefont {Amico}, \citenamefont {Burchianti}, \citenamefont {Fort},
  \citenamefont {Inguscio}, \citenamefont {Zaccanti},\ and\ \citenamefont
  {Roati}}]{Scazza2017}%
  \BibitemOpen
  \bibfield  {author} {\bibinfo {author} {\bibfnamefont {F.}~\bibnamefont
  {Scazza}}, \bibinfo {author} {\bibfnamefont {G.}~\bibnamefont {Valtolina}},
  \bibinfo {author} {\bibfnamefont {P.}~\bibnamefont {Massignan}}, \bibinfo
  {author} {\bibfnamefont {A.}~\bibnamefont {Recati}}, \bibinfo {author}
  {\bibfnamefont {A.}~\bibnamefont {Amico}}, \bibinfo {author} {\bibfnamefont
  {A.}~\bibnamefont {Burchianti}}, \bibinfo {author} {\bibfnamefont
  {C.}~\bibnamefont {Fort}}, \bibinfo {author} {\bibfnamefont {M.}~\bibnamefont
  {Inguscio}}, \bibinfo {author} {\bibfnamefont {M.}~\bibnamefont {Zaccanti}},
  \ and\ \bibinfo {author} {\bibfnamefont {G.}~\bibnamefont {Roati}},\
  }\bibfield  {title} {\bibinfo {title} {\emph {{Repulsive Fermi Polarons in a
  Resonant Mixture of Ultracold 6-Li Atoms}}},\ }\href {\doibase
  10.1103/PhysRevLett.118.083602} {\bibfield  {journal} {\bibinfo  {journal}
  {Phys. Rev. Lett.}\ }\textbf {\bibinfo {volume} {118}},\ \bibinfo {pages}
  {083602} (\bibinfo {year} {2017})}\BibitemShut {NoStop}%
\bibitem [{\citenamefont {Yan}\ \emph {et~al.}(2019)\citenamefont {Yan},
  \citenamefont {Patel}, \citenamefont {Mukherjee}, \citenamefont {Fletcher},
  \citenamefont {Struck},\ and\ \citenamefont {Zwierlein}}]{Yan2019}%
  \BibitemOpen
  \bibfield  {author} {\bibinfo {author} {\bibfnamefont {Z.}~\bibnamefont
  {Yan}}, \bibinfo {author} {\bibfnamefont {P.~B.}\ \bibnamefont {Patel}},
  \bibinfo {author} {\bibfnamefont {B.}~\bibnamefont {Mukherjee}}, \bibinfo
  {author} {\bibfnamefont {R.~J.}\ \bibnamefont {Fletcher}}, \bibinfo {author}
  {\bibfnamefont {J.}~\bibnamefont {Struck}}, \ and\ \bibinfo {author}
  {\bibfnamefont {M.~W.}\ \bibnamefont {Zwierlein}},\ }\bibfield  {title}
  {\bibinfo {title} {\emph {{Boiling a Unitary Fermi Liquid}}},\ }\href
  {\doibase 10.1103/PhysRevLett.122.093401} {\bibfield  {journal} {\bibinfo
  {journal} {Phys. Rev. Lett.}\ }\textbf {\bibinfo {volume} {122}},\ \bibinfo
  {pages} {093401} (\bibinfo {year} {2019})}\BibitemShut {NoStop}%
\bibitem [{\citenamefont {Darkwah~Oppong}\ \emph {et~al.}(2019)\citenamefont
  {Darkwah~Oppong}, \citenamefont {Riegger}, \citenamefont {Bettermann},
  \citenamefont {H\"ofer}, \citenamefont {Levinsen}, \citenamefont {Parish},
  \citenamefont {Bloch},\ and\ \citenamefont {F\"olling}}]{Oppong2019}%
  \BibitemOpen
  \bibfield  {author} {\bibinfo {author} {\bibfnamefont {N.}~\bibnamefont
  {Darkwah~Oppong}}, \bibinfo {author} {\bibfnamefont {L.}~\bibnamefont
  {Riegger}}, \bibinfo {author} {\bibfnamefont {O.}~\bibnamefont {Bettermann}},
  \bibinfo {author} {\bibfnamefont {M.}~\bibnamefont {H\"ofer}}, \bibinfo
  {author} {\bibfnamefont {J.}~\bibnamefont {Levinsen}}, \bibinfo {author}
  {\bibfnamefont {M.~M.}\ \bibnamefont {Parish}}, \bibinfo {author}
  {\bibfnamefont {I.}~\bibnamefont {Bloch}}, \ and\ \bibinfo {author}
  {\bibfnamefont {S.}~\bibnamefont {F\"olling}},\ }\bibfield  {title} {\bibinfo
  {title} {\emph {Observation of Coherent Multiorbital Polarons in a
  Two-Dimensional Fermi Gas}},\ }\href {\doibase
  10.1103/PhysRevLett.122.193604} {\bibfield  {journal} {\bibinfo  {journal}
  {Phys. Rev. Lett.}\ }\textbf {\bibinfo {volume} {122}},\ \bibinfo {pages}
  {193604} (\bibinfo {year} {2019})}\BibitemShut {NoStop}%
\bibitem [{\citenamefont {Ness}\ \emph {et~al.}(2020)\citenamefont {Ness},
  \citenamefont {Shkedrov}, \citenamefont {Florshaim}, \citenamefont {Diessel},
  \citenamefont {von Milczewski}, \citenamefont {Schmidt},\ and\ \citenamefont
  {Sagi}}]{Ness2020}%
  \BibitemOpen
  \bibfield  {author} {\bibinfo {author} {\bibfnamefont {G.}~\bibnamefont
  {Ness}}, \bibinfo {author} {\bibfnamefont {C.}~\bibnamefont {Shkedrov}},
  \bibinfo {author} {\bibfnamefont {Y.}~\bibnamefont {Florshaim}}, \bibinfo
  {author} {\bibfnamefont {O.~K.}\ \bibnamefont {Diessel}}, \bibinfo {author}
  {\bibfnamefont {J.}~\bibnamefont {von Milczewski}}, \bibinfo {author}
  {\bibfnamefont {R.}~\bibnamefont {Schmidt}}, \ and\ \bibinfo {author}
  {\bibfnamefont {Y.}~\bibnamefont {Sagi}},\ }\bibfield  {title} {\bibinfo
  {title} {\emph {Observation of a Smooth Polaron-Molecule Transition in a
  Degenerate Fermi Gas}},\ }\href {\doibase 10.1103/PhysRevX.10.041019}
  {\bibfield  {journal} {\bibinfo  {journal} {Phys. Rev. X}\ }\textbf {\bibinfo
  {volume} {10}},\ \bibinfo {pages} {041019} (\bibinfo {year}
  {2020})}\BibitemShut {NoStop}%
\bibitem [{\citenamefont {Catani}\ \emph {et~al.}(2012)\citenamefont {Catani},
  \citenamefont {Lamporesi}, \citenamefont {Naik}, \citenamefont {Gring},
  \citenamefont {Inguscio}, \citenamefont {Minardi}, \citenamefont {Kantian},\
  and\ \citenamefont {Giamarchi}}]{Catani2012}%
  \BibitemOpen
  \bibfield  {author} {\bibinfo {author} {\bibfnamefont {J.}~\bibnamefont
  {Catani}}, \bibinfo {author} {\bibfnamefont {G.}~\bibnamefont {Lamporesi}},
  \bibinfo {author} {\bibfnamefont {D.}~\bibnamefont {Naik}}, \bibinfo {author}
  {\bibfnamefont {M.}~\bibnamefont {Gring}}, \bibinfo {author} {\bibfnamefont
  {M.}~\bibnamefont {Inguscio}}, \bibinfo {author} {\bibfnamefont
  {F.}~\bibnamefont {Minardi}}, \bibinfo {author} {\bibfnamefont
  {A.}~\bibnamefont {Kantian}}, \ and\ \bibinfo {author} {\bibfnamefont
  {T.}~\bibnamefont {Giamarchi}},\ }\bibfield  {title} {\bibinfo {title} {\emph
  {{Quantum dynamics of impurities in a one-dimensional Bose gas}}},\ }\href
  {\doibase 10.1103/PhysRevA.85.023623} {\bibfield  {journal} {\bibinfo
  {journal} {Phys. Rev. A}\ }\textbf {\bibinfo {volume} {85}},\ \bibinfo
  {pages} {023623} (\bibinfo {year} {2012})}\BibitemShut {NoStop}%
\bibitem [{\citenamefont {Hu}\ \emph {et~al.}(2016)\citenamefont {Hu},
  \citenamefont {de~Graaff}, \citenamefont {Kedar}, \citenamefont {Corson},
  \citenamefont {Cornell},\ and\ \citenamefont {Jin}}]{Hu2016}%
  \BibitemOpen
  \bibfield  {author} {\bibinfo {author} {\bibfnamefont {M.-G.}\ \bibnamefont
  {Hu}}, \bibinfo {author} {\bibfnamefont {M.~J.}\ \bibnamefont {de~Graaff}},
  \bibinfo {author} {\bibfnamefont {D.}~\bibnamefont {Kedar}}, \bibinfo
  {author} {\bibfnamefont {J.~P.}\ \bibnamefont {Corson}}, \bibinfo {author}
  {\bibfnamefont {E.~A.}\ \bibnamefont {Cornell}}, \ and\ \bibinfo {author}
  {\bibfnamefont {D.~S.}\ \bibnamefont {Jin}},\ }\bibfield  {title} {\bibinfo
  {title} {\emph {{Bose Polarons in the Strongly Interacting Regime}}},\ }\href
  {\doibase 10.1103/PhysRevLett.117.055301} {\bibfield  {journal} {\bibinfo
  {journal} {Phys. Rev. Lett.}\ }\textbf {\bibinfo {volume} {117}},\ \bibinfo
  {pages} {055301} (\bibinfo {year} {2016})}\BibitemShut {NoStop}%
\bibitem [{\citenamefont {J{\o}rgensen}\ \emph {et~al.}(2016)\citenamefont
  {J{\o}rgensen}, \citenamefont {Wacker}, \citenamefont {Skalmstang},
  \citenamefont {Parish}, \citenamefont {Levinsen}, \citenamefont
  {Christensen}, \citenamefont {Bruun},\ and\ \citenamefont
  {Arlt}}]{Jorgensen2016}%
  \BibitemOpen
  \bibfield  {author} {\bibinfo {author} {\bibfnamefont {N.~B.}\ \bibnamefont
  {J{\o}rgensen}}, \bibinfo {author} {\bibfnamefont {L.}~\bibnamefont
  {Wacker}}, \bibinfo {author} {\bibfnamefont {K.~T.}\ \bibnamefont
  {Skalmstang}}, \bibinfo {author} {\bibfnamefont {M.~M.}\ \bibnamefont
  {Parish}}, \bibinfo {author} {\bibfnamefont {J.}~\bibnamefont {Levinsen}},
  \bibinfo {author} {\bibfnamefont {R.~S.}\ \bibnamefont {Christensen}},
  \bibinfo {author} {\bibfnamefont {G.~M.}\ \bibnamefont {Bruun}}, \ and\
  \bibinfo {author} {\bibfnamefont {J.~J.}\ \bibnamefont {Arlt}},\ }\bibfield
  {title} {\bibinfo {title} {\emph {{Observation of Attractive and Repulsive
  Polarons in a Bose-Einstein Condensate}}},\ }\href {\doibase
  10.1103/PhysRevLett.117.055302} {\bibfield  {journal} {\bibinfo  {journal}
  {Phys. Rev. Lett.}\ }\textbf {\bibinfo {volume} {117}},\ \bibinfo {pages}
  {055302} (\bibinfo {year} {2016})}\BibitemShut {NoStop}%
\bibitem [{\citenamefont {Camargo}\ \emph {et~al.}(2018)\citenamefont
  {Camargo}, \citenamefont {Schmidt}, \citenamefont {Whalen}, \citenamefont
  {Ding}, \citenamefont {Woehl}, \citenamefont {Yoshida}, \citenamefont
  {Burgd{\"{o}}rfer}, \citenamefont {Dunning}, \citenamefont {Sadeghpour},
  \citenamefont {Demler},\ and\ \citenamefont {Killian}}]{Camargo2018}%
  \BibitemOpen
  \bibfield  {author} {\bibinfo {author} {\bibfnamefont {F.}~\bibnamefont
  {Camargo}}, \bibinfo {author} {\bibfnamefont {R.}~\bibnamefont {Schmidt}},
  \bibinfo {author} {\bibfnamefont {J.~D.}\ \bibnamefont {Whalen}}, \bibinfo
  {author} {\bibfnamefont {R.}~\bibnamefont {Ding}}, \bibinfo {author}
  {\bibfnamefont {G.}~\bibnamefont {Woehl}}, \bibinfo {author} {\bibfnamefont
  {S.}~\bibnamefont {Yoshida}}, \bibinfo {author} {\bibfnamefont
  {J.}~\bibnamefont {Burgd{\"{o}}rfer}}, \bibinfo {author} {\bibfnamefont
  {F.~B.}\ \bibnamefont {Dunning}}, \bibinfo {author} {\bibfnamefont {H.~R.}\
  \bibnamefont {Sadeghpour}}, \bibinfo {author} {\bibfnamefont
  {E.}~\bibnamefont {Demler}}, \ and\ \bibinfo {author} {\bibfnamefont {T.~C.}\
  \bibnamefont {Killian}},\ }\bibfield  {title} {\bibinfo {title} {\emph
  {{Creation of Rydberg Polarons in a Bose Gas}}},\ }\href {\doibase
  10.1103/PhysRevLett.120.083401} {\bibfield  {journal} {\bibinfo  {journal}
  {Phys. Rev. Lett.}\ }\textbf {\bibinfo {volume} {120}},\ \bibinfo {pages}
  {083401} (\bibinfo {year} {2018})}\BibitemShut {NoStop}%
\bibitem [{\citenamefont {Yan}\ \emph {et~al.}(2020)\citenamefont {Yan},
  \citenamefont {Ni}, \citenamefont {Robens},\ and\ \citenamefont
  {Zwierlein}}]{Yan2019u}%
  \BibitemOpen
  \bibfield  {author} {\bibinfo {author} {\bibfnamefont {Z.~Z.}\ \bibnamefont
  {Yan}}, \bibinfo {author} {\bibfnamefont {Y.}~\bibnamefont {Ni}}, \bibinfo
  {author} {\bibfnamefont {C.}~\bibnamefont {Robens}}, \ and\ \bibinfo {author}
  {\bibfnamefont {M.~W.}\ \bibnamefont {Zwierlein}},\ }\bibfield  {title}
  {\bibinfo {title} {\emph {Bose polarons near quantum criticality}},\ }\href
  {\doibase 10.1126/science.aax5850} {\bibfield  {journal} {\bibinfo  {journal}
  {Science}\ }\textbf {\bibinfo {volume} {368}},\ \bibinfo {pages} {190}
  (\bibinfo {year} {2020})}\BibitemShut {NoStop}%
\bibitem [{\citenamefont {Sidler}\ \emph {et~al.}(2017)\citenamefont {Sidler},
  \citenamefont {Back}, \citenamefont {Cotlet}, \citenamefont {Srivastava},
  \citenamefont {Fink}, \citenamefont {Kroner}, \citenamefont {Demler},\ and\
  \citenamefont {Imamoglu}}]{Sidler2017}%
  \BibitemOpen
  \bibfield  {author} {\bibinfo {author} {\bibfnamefont {M.}~\bibnamefont
  {Sidler}}, \bibinfo {author} {\bibfnamefont {P.}~\bibnamefont {Back}},
  \bibinfo {author} {\bibfnamefont {O.}~\bibnamefont {Cotlet}}, \bibinfo
  {author} {\bibfnamefont {A.}~\bibnamefont {Srivastava}}, \bibinfo {author}
  {\bibfnamefont {T.}~\bibnamefont {Fink}}, \bibinfo {author} {\bibfnamefont
  {M.}~\bibnamefont {Kroner}}, \bibinfo {author} {\bibfnamefont
  {E.}~\bibnamefont {Demler}}, \ and\ \bibinfo {author} {\bibfnamefont
  {A.}~\bibnamefont {Imamoglu}},\ }\bibfield  {title} {\bibinfo {title} {\emph
  {{Fermi polaron-polaritons in charge-tunable atomically thin
  semiconductors}}},\ }\href {\doibase 10.1038/nphys3949} {\bibfield  {journal}
  {\bibinfo  {journal} {Nat. Phys.}\ }\textbf {\bibinfo {volume} {13}},\
  \bibinfo {pages} {255} (\bibinfo {year} {2017})}\BibitemShut {NoStop}%
\bibitem [{\citenamefont {Chevy}(2006)}]{Chevy2006upd}%
  \BibitemOpen
  \bibfield  {author} {\bibinfo {author} {\bibfnamefont {F.}~\bibnamefont
  {Chevy}},\ }\bibfield  {title} {\bibinfo {title} {\emph {Universal phase
  diagram of a strongly interacting Fermi gas with unbalanced spin
  populations}},\ }\href {\doibase 10.1103/PhysRevA.74.063628} {\bibfield
  {journal} {\bibinfo  {journal} {Phys. Rev. A}\ }\textbf {\bibinfo {volume}
  {74}},\ \bibinfo {pages} {063628} (\bibinfo {year} {2006})}\BibitemShut
  {NoStop}%
\bibitem [{\citenamefont {Combescot}\ \emph {et~al.}(2007)\citenamefont
  {Combescot}, \citenamefont {Recati}, \citenamefont {Lobo},\ and\
  \citenamefont {Chevy}}]{Combescot2007}%
  \BibitemOpen
  \bibfield  {author} {\bibinfo {author} {\bibfnamefont {R.}~\bibnamefont
  {Combescot}}, \bibinfo {author} {\bibfnamefont {A.}~\bibnamefont {Recati}},
  \bibinfo {author} {\bibfnamefont {C.}~\bibnamefont {Lobo}}, \ and\ \bibinfo
  {author} {\bibfnamefont {F.}~\bibnamefont {Chevy}},\ }\bibfield  {title}
  {\bibinfo {title} {\emph {{Normal State of Highly Polarized Fermi Gases:
  Simple Many-Body Approaches}}},\ }\href
  {http://dx.doi.org/10.1103/PhysRevLett.98.180402} {\bibfield  {journal}
  {\bibinfo  {journal} {Phys. Rev. Lett.}\ }\textbf {\bibinfo {volume} {98}},\
  \bibinfo {pages} {180402} (\bibinfo {year} {2007})}\BibitemShut {NoStop}%
\bibitem [{\citenamefont {Prokof'ev}\ and\ \citenamefont
  {Svistunov}(2008)}]{Prokofev2008}%
  \BibitemOpen
  \bibfield  {author} {\bibinfo {author} {\bibfnamefont {N.}~\bibnamefont
  {Prokof'ev}}\ and\ \bibinfo {author} {\bibfnamefont {B.}~\bibnamefont
  {Svistunov}},\ }\bibfield  {title} {\bibinfo {title} {\emph {Fermi-polaron
  problem: Diagrammatic Monte Carlo method for divergent sign-alternating
  series}},\ }\href {\doibase 10.1103/PhysRevB.77.020408} {\bibfield  {journal}
  {\bibinfo  {journal} {Phys. Rev. B}\ }\textbf {\bibinfo {volume} {77}},\
  \bibinfo {pages} {020408} (\bibinfo {year} {2008})}\BibitemShut {NoStop}%
\bibitem [{\citenamefont {Combescot}\ and\ \citenamefont
  {Giraud}(2008)}]{Combescot2008}%
  \BibitemOpen
  \bibfield  {author} {\bibinfo {author} {\bibfnamefont {R.}~\bibnamefont
  {Combescot}}\ and\ \bibinfo {author} {\bibfnamefont {S.}~\bibnamefont
  {Giraud}},\ }\bibfield  {title} {\bibinfo {title} {\emph {{Normal state of
  highly polarized fermi gases: Full many-body treatment}}},\ }\href {\doibase
  10.1103/PhysRevLett.101.050404} {\bibfield  {journal} {\bibinfo  {journal}
  {Phys. Rev. Lett.}\ }\textbf {\bibinfo {volume} {101}},\ \bibinfo {pages}
  {050404} (\bibinfo {year} {2008})}\BibitemShut {NoStop}%
\bibitem [{\citenamefont {Punk}\ \emph {et~al.}(2009)\citenamefont {Punk},
  \citenamefont {Dumitrescu},\ and\ \citenamefont {Zwerger}}]{Punk2009}%
  \BibitemOpen
  \bibfield  {author} {\bibinfo {author} {\bibfnamefont {M.}~\bibnamefont
  {Punk}}, \bibinfo {author} {\bibfnamefont {P.~T.}\ \bibnamefont
  {Dumitrescu}}, \ and\ \bibinfo {author} {\bibfnamefont {W.}~\bibnamefont
  {Zwerger}},\ }\bibfield  {title} {\bibinfo {title} {\emph
  {{Polaron-to-molecule transition in a strongly imbalanced Fermi gas}}},\
  }\href {\doibase 10.1103/PhysRevA.80.053605} {\bibfield  {journal} {\bibinfo
  {journal} {Phys. Rev. A}\ }\textbf {\bibinfo {volume} {80}},\ \bibinfo
  {pages} {053605} (\bibinfo {year} {2009})}\BibitemShut {NoStop}%
\bibitem [{\citenamefont {Mathy}\ \emph {et~al.}(2011)\citenamefont {Mathy},
  \citenamefont {Parish},\ and\ \citenamefont {Huse}}]{Mathy2011}%
  \BibitemOpen
  \bibfield  {author} {\bibinfo {author} {\bibfnamefont {C.~J.~M.}\
  \bibnamefont {Mathy}}, \bibinfo {author} {\bibfnamefont {M.~M.}\ \bibnamefont
  {Parish}}, \ and\ \bibinfo {author} {\bibfnamefont {D.~A.}\ \bibnamefont
  {Huse}},\ }\bibfield  {title} {\bibinfo {title} {\emph {{Trimers, Molecules,
  and Polarons in Mass-Imbalanced Atomic Fermi Gases}}},\ }\href {\doibase
  10.1103/PhysRevLett.106.166404} {\bibfield  {journal} {\bibinfo  {journal}
  {Phys. Rev. Lett.}\ }\textbf {\bibinfo {volume} {106}},\ \bibinfo {pages}
  {166404} (\bibinfo {year} {2011})}\BibitemShut {NoStop}%
\bibitem [{\citenamefont {Schmidt}\ and\ \citenamefont
  {Enss}(2011)}]{Schmidt2011}%
  \BibitemOpen
  \bibfield  {author} {\bibinfo {author} {\bibfnamefont {R.}~\bibnamefont
  {Schmidt}}\ and\ \bibinfo {author} {\bibfnamefont {T.}~\bibnamefont {Enss}},\
  }\bibfield  {title} {\bibinfo {title} {\emph {{Excitation spectra and rf
  response near the polaron-to-molecule transition from the functional
  renormalization group}}},\ }\href {\doibase 10.1103/PhysRevA.83.063620}
  {\bibfield  {journal} {\bibinfo  {journal} {Phys. Rev. A}\ }\textbf {\bibinfo
  {volume} {83}},\ \bibinfo {pages} {063620} (\bibinfo {year}
  {2011})}\BibitemShut {NoStop}%
\bibitem [{\citenamefont {Trefzger}\ and\ \citenamefont
  {Castin}(2012)}]{Trefzger2012}%
  \BibitemOpen
  \bibfield  {author} {\bibinfo {author} {\bibfnamefont {C.}~\bibnamefont
  {Trefzger}}\ and\ \bibinfo {author} {\bibfnamefont {Y.}~\bibnamefont
  {Castin}},\ }\bibfield  {title} {\bibinfo {title} {\emph {Impurity in a Fermi
  sea on a narrow Feshbach resonance: A variational study of the polaronic and
  dimeronic branches}},\ }\href {\doibase 10.1103/PhysRevA.85.053612}
  {\bibfield  {journal} {\bibinfo  {journal} {Phys. Rev. A}\ }\textbf {\bibinfo
  {volume} {85}},\ \bibinfo {pages} {053612} (\bibinfo {year}
  {2012})}\BibitemShut {NoStop}%
\bibitem [{\citenamefont {Levinsen}\ and\ \citenamefont
  {Parish}(2015)}]{LevinsenResonant2DGases}%
  \BibitemOpen
  \bibfield  {author} {\bibinfo {author} {\bibfnamefont {J.}~\bibnamefont
  {Levinsen}}\ and\ \bibinfo {author} {\bibfnamefont {M.~M.}\ \bibnamefont
  {Parish}},\ }\bibfield  {title} {\bibinfo {title} {\emph {{Strongly
  interacting two-dimensional Fermi gases}}},\ }\href
  {https://doi.org/10.1142/9789814667746_0001} {\bibfield  {journal} {\bibinfo
  {journal} {Annu. Rev. Cold Atoms Mol.}\ }\textbf {\bibinfo {volume} {3}},\
  \bibinfo {pages} {1} (\bibinfo {year} {2015})}\BibitemShut {NoStop}%
\bibitem [{\citenamefont {Cui}\ and\ \citenamefont {Zhai}(2010)}]{Cui2010}%
  \BibitemOpen
  \bibfield  {author} {\bibinfo {author} {\bibfnamefont {X.}~\bibnamefont
  {Cui}}\ and\ \bibinfo {author} {\bibfnamefont {H.}~\bibnamefont {Zhai}},\
  }\bibfield  {title} {\bibinfo {title} {\emph {{Stability of a fully
  magnetized ferromagnetic state in repulsively interacting ultracold Fermi
  gases}}},\ }\href {http://dx.doi.org/10.1103/PhysRevA.81.041602} {\bibfield
  {journal} {\bibinfo  {journal} {Phys. Rev. A}\ }\textbf {\bibinfo {volume}
  {81}},\ \bibinfo {pages} {041602} (\bibinfo {year} {2010})}\BibitemShut
  {NoStop}%
\bibitem [{\citenamefont {Pilati}\ \emph {et~al.}(2010)\citenamefont {Pilati},
  \citenamefont {Bertaina}, \citenamefont {Giorgini},\ and\ \citenamefont
  {Troyer}}]{Pilati2010}%
  \BibitemOpen
  \bibfield  {author} {\bibinfo {author} {\bibfnamefont {S.}~\bibnamefont
  {Pilati}}, \bibinfo {author} {\bibfnamefont {G.}~\bibnamefont {Bertaina}},
  \bibinfo {author} {\bibfnamefont {S.}~\bibnamefont {Giorgini}}, \ and\
  \bibinfo {author} {\bibfnamefont {M.}~\bibnamefont {Troyer}},\ }\bibfield
  {title} {\bibinfo {title} {\emph {{Itinerant Ferromagnetism of a Repulsive
  Atomic Fermi Gas: A Quantum Monte Carlo Study}}},\ }\href {\doibase
  10.1103/PhysRevLett.105.030405} {\bibfield  {journal} {\bibinfo  {journal}
  {Phys. Rev. Lett.}\ }\textbf {\bibinfo {volume} {105}},\ \bibinfo {pages}
  {030405} (\bibinfo {year} {2010})}\BibitemShut {NoStop}%
\bibitem [{\citenamefont {Massignan}\ and\ \citenamefont
  {Bruun}(2011)}]{Massignan2011}%
  \BibitemOpen
  \bibfield  {author} {\bibinfo {author} {\bibfnamefont {P.}~\bibnamefont
  {Massignan}}\ and\ \bibinfo {author} {\bibfnamefont {G.~M.}\ \bibnamefont
  {Bruun}},\ }\bibfield  {title} {\bibinfo {title} {\emph {{Repulsive polarons
  and itinerant ferromagnetism in strongly polarized Fermi gases}}},\ }\href
  {\doibase 10.1140/epjd/e2011-20084-5} {\bibfield  {journal} {\bibinfo
  {journal} {Eur. Phys. J. D}\ }\textbf {\bibinfo {volume} {65}},\ \bibinfo
  {pages} {83} (\bibinfo {year} {2011})}\BibitemShut {NoStop}%
\bibitem [{\citenamefont {Goulko}\ \emph {et~al.}(2016)\citenamefont {Goulko},
  \citenamefont {Mishchenko}, \citenamefont {Prokof'ev},\ and\ \citenamefont
  {Svistunov}}]{Goulko2016}%
  \BibitemOpen
  \bibfield  {author} {\bibinfo {author} {\bibfnamefont {O.}~\bibnamefont
  {Goulko}}, \bibinfo {author} {\bibfnamefont {A.~S.}\ \bibnamefont
  {Mishchenko}}, \bibinfo {author} {\bibfnamefont {N.}~\bibnamefont
  {Prokof'ev}}, \ and\ \bibinfo {author} {\bibfnamefont {B.}~\bibnamefont
  {Svistunov}},\ }\bibfield  {title} {\bibinfo {title} {\emph {{Dark continuum
  in the spectral function of the resonant Fermi polaron}}},\ }\href {\doibase
  10.1103/PhysRevA.94.051605} {\bibfield  {journal} {\bibinfo  {journal} {Phys.
  Rev. A}\ }\textbf {\bibinfo {volume} {94}},\ \bibinfo {pages} {051605}
  (\bibinfo {year} {2016})}\BibitemShut {NoStop}%
\bibitem [{\citenamefont {Tajima}\ and\ \citenamefont
  {Uchino}(2018)}]{Tajima2018}%
  \BibitemOpen
  \bibfield  {author} {\bibinfo {author} {\bibfnamefont {H.}~\bibnamefont
  {Tajima}}\ and\ \bibinfo {author} {\bibfnamefont {S.}~\bibnamefont
  {Uchino}},\ }\bibfield  {title} {\bibinfo {title} {\emph {Many Fermi polarons
  at nonzero temperature}},\ }\href {\doibase 10.1088/1367-2630/aad1e7}
  {\bibfield  {journal} {\bibinfo  {journal} {New J. Phys.}\ }\textbf {\bibinfo
  {volume} {20}},\ \bibinfo {pages} {073048} (\bibinfo {year}
  {2018})}\BibitemShut {NoStop}%
\bibitem [{\citenamefont {Mulkerin}\ \emph {et~al.}(2019)\citenamefont
  {Mulkerin}, \citenamefont {Liu},\ and\ \citenamefont {Hu}}]{Mulkerin2019}%
  \BibitemOpen
  \bibfield  {author} {\bibinfo {author} {\bibfnamefont {B.~C.}\ \bibnamefont
  {Mulkerin}}, \bibinfo {author} {\bibfnamefont {X.-J.}\ \bibnamefont {Liu}}, \
  and\ \bibinfo {author} {\bibfnamefont {H.}~\bibnamefont {Hu}},\ }\bibfield
  {title} {\bibinfo {title} {\emph {{Breakdown of the Fermi polaron description
  near Fermi degeneracy at unitarity}}},\ }\href {\doibase
  10.1016/j.aop.2019.04.018} {\bibfield  {journal} {\bibinfo  {journal} {Ann.
  Phys. (N.~Y.)}\ }\textbf {\bibinfo {volume} {407}},\ \bibinfo {pages} {29}
  (\bibinfo {year} {2019})}\BibitemShut {NoStop}%
\bibitem [{\citenamefont {Pekker}\ \emph {et~al.}(2011)\citenamefont {Pekker},
  \citenamefont {Babadi}, \citenamefont {Sensarma}, \citenamefont {Zinner},
  \citenamefont {Pollet}, \citenamefont {Zwierlein},\ and\ \citenamefont
  {Demler}}]{Pekker2011cbp}%
  \BibitemOpen
  \bibfield  {author} {\bibinfo {author} {\bibfnamefont {D.}~\bibnamefont
  {Pekker}}, \bibinfo {author} {\bibfnamefont {M.}~\bibnamefont {Babadi}},
  \bibinfo {author} {\bibfnamefont {R.}~\bibnamefont {Sensarma}}, \bibinfo
  {author} {\bibfnamefont {N.}~\bibnamefont {Zinner}}, \bibinfo {author}
  {\bibfnamefont {L.}~\bibnamefont {Pollet}}, \bibinfo {author} {\bibfnamefont
  {M.~W.}\ \bibnamefont {Zwierlein}}, \ and\ \bibinfo {author} {\bibfnamefont
  {E.}~\bibnamefont {Demler}},\ }\bibfield  {title} {\bibinfo {title} {\emph
  {Competition between Pairing and Ferromagnetic Instabilities in Ultracold
  Fermi Gases near Feshbach Resonances}},\ }\href {\doibase
  10.1103/PhysRevLett.106.050402} {\bibfield  {journal} {\bibinfo  {journal}
  {Phys. Rev. Lett.}\ }\textbf {\bibinfo {volume} {106}},\ \bibinfo {pages}
  {050402} (\bibinfo {year} {2011})}\BibitemShut {NoStop}%
\bibitem [{\citenamefont {Sanner}\ \emph {et~al.}(2012)\citenamefont {Sanner},
  \citenamefont {Su}, \citenamefont {Huang}, \citenamefont {Keshet},
  \citenamefont {Gillen},\ and\ \citenamefont {Ketterle}}]{Sanner2012}%
  \BibitemOpen
  \bibfield  {author} {\bibinfo {author} {\bibfnamefont {C.}~\bibnamefont
  {Sanner}}, \bibinfo {author} {\bibfnamefont {E.~J.}\ \bibnamefont {Su}},
  \bibinfo {author} {\bibfnamefont {W.}~\bibnamefont {Huang}}, \bibinfo
  {author} {\bibfnamefont {A.}~\bibnamefont {Keshet}}, \bibinfo {author}
  {\bibfnamefont {J.}~\bibnamefont {Gillen}}, \ and\ \bibinfo {author}
  {\bibfnamefont {W.}~\bibnamefont {Ketterle}},\ }\bibfield  {title} {\bibinfo
  {title} {\emph {Correlations and Pair Formation in a Repulsively Interacting
  Fermi Gas}},\ }\href {\doibase 10.1103/PhysRevLett.108.240404} {\bibfield
  {journal} {\bibinfo  {journal} {Phys. Rev. Lett.}\ }\textbf {\bibinfo
  {volume} {108}},\ \bibinfo {pages} {240404} (\bibinfo {year}
  {2012})}\BibitemShut {NoStop}%
\bibitem [{\citenamefont {Valtolina}\ \emph {et~al.}(2017)\citenamefont
  {Valtolina}, \citenamefont {Scazza}, \citenamefont {Amico}, \citenamefont
  {Burchianti}, \citenamefont {Recati}, \citenamefont {Enss}, \citenamefont
  {Inguscio}, \citenamefont {Zaccanti},\ and\ \citenamefont
  {Roati}}]{Valtolina2017}%
  \BibitemOpen
  \bibfield  {author} {\bibinfo {author} {\bibfnamefont {G.}~\bibnamefont
  {Valtolina}}, \bibinfo {author} {\bibfnamefont {F.}~\bibnamefont {Scazza}},
  \bibinfo {author} {\bibfnamefont {A.}~\bibnamefont {Amico}}, \bibinfo
  {author} {\bibfnamefont {A.}~\bibnamefont {Burchianti}}, \bibinfo {author}
  {\bibfnamefont {A.}~\bibnamefont {Recati}}, \bibinfo {author} {\bibfnamefont
  {T.}~\bibnamefont {Enss}}, \bibinfo {author} {\bibfnamefont {M.}~\bibnamefont
  {Inguscio}}, \bibinfo {author} {\bibfnamefont {M.}~\bibnamefont {Zaccanti}},
  \ and\ \bibinfo {author} {\bibfnamefont {G.}~\bibnamefont {Roati}},\
  }\bibfield  {title} {\bibinfo {title} {\emph {Exploring the ferromagnetic
  behaviour of a repulsive Fermi gas through spin dynamics}},\ }\href {\doibase
  10.1038/nphys4108} {\bibfield  {journal} {\bibinfo  {journal} {Nature Phys.}\
  }\textbf {\bibinfo {volume} {13}},\ \bibinfo {pages} {704} (\bibinfo {year}
  {2017})}\BibitemShut {NoStop}%
\bibitem [{\citenamefont {Amico}\ \emph {et~al.}(2018)\citenamefont {Amico},
  \citenamefont {Scazza}, \citenamefont {Valtolina}, \citenamefont {Tavares},
  \citenamefont {Ketterle}, \citenamefont {Inguscio}, \citenamefont {Roati},\
  and\ \citenamefont {Zaccanti}}]{Amico2018}%
  \BibitemOpen
  \bibfield  {author} {\bibinfo {author} {\bibfnamefont {A.}~\bibnamefont
  {Amico}}, \bibinfo {author} {\bibfnamefont {F.}~\bibnamefont {Scazza}},
  \bibinfo {author} {\bibfnamefont {G.}~\bibnamefont {Valtolina}}, \bibinfo
  {author} {\bibfnamefont {P.~E.~S.}\ \bibnamefont {Tavares}}, \bibinfo
  {author} {\bibfnamefont {W.}~\bibnamefont {Ketterle}}, \bibinfo {author}
  {\bibfnamefont {M.}~\bibnamefont {Inguscio}}, \bibinfo {author}
  {\bibfnamefont {G.}~\bibnamefont {Roati}}, \ and\ \bibinfo {author}
  {\bibfnamefont {M.}~\bibnamefont {Zaccanti}},\ }\bibfield  {title} {\bibinfo
  {title} {\emph {Time-Resolved Observation of Competing Attractive and
  Repulsive Short-Range Correlations in Strongly Interacting Fermi Gases}},\
  }\href {\doibase 10.1103/PhysRevLett.121.253602} {\bibfield  {journal}
  {\bibinfo  {journal} {Phys. Rev. Lett.}\ }\textbf {\bibinfo {volume} {121}},\
  \bibinfo {pages} {253602} (\bibinfo {year} {2018})}\BibitemShut {NoStop}%
\bibitem [{\citenamefont {Scazza}\ \emph {et~al.}(2020)\citenamefont {Scazza},
  \citenamefont {Valtolina}, \citenamefont {Amico}, \citenamefont {Tavares},
  \citenamefont {Inguscio}, \citenamefont {Ketterle}, \citenamefont {Roati},\
  and\ \citenamefont {Zaccanti}}]{Scazza2020}%
  \BibitemOpen
  \bibfield  {author} {\bibinfo {author} {\bibfnamefont {F.}~\bibnamefont
  {Scazza}}, \bibinfo {author} {\bibfnamefont {G.}~\bibnamefont {Valtolina}},
  \bibinfo {author} {\bibfnamefont {A.}~\bibnamefont {Amico}}, \bibinfo
  {author} {\bibfnamefont {P.~E.~S.}\ \bibnamefont {Tavares}}, \bibinfo
  {author} {\bibfnamefont {M.}~\bibnamefont {Inguscio}}, \bibinfo {author}
  {\bibfnamefont {W.}~\bibnamefont {Ketterle}}, \bibinfo {author}
  {\bibfnamefont {G.}~\bibnamefont {Roati}}, \ and\ \bibinfo {author}
  {\bibfnamefont {M.}~\bibnamefont {Zaccanti}},\ }\bibfield  {title} {\bibinfo
  {title} {\emph {Exploring emergent heterogeneous phases in strongly repulsive
  Fermi gases}},\ }\href {\doibase 10.1103/PhysRevA.101.013603} {\bibfield
  {journal} {\bibinfo  {journal} {Phys. Rev. A}\ }\textbf {\bibinfo {volume}
  {101}},\ \bibinfo {pages} {013603} (\bibinfo {year} {2020})}\BibitemShut
  {NoStop}%
\bibitem [{\citenamefont {Liu}\ \emph {et~al.}(2019)\citenamefont {Liu},
  \citenamefont {Levinsen},\ and\ \citenamefont {Parish}}]{Liu2019}%
  \BibitemOpen
  \bibfield  {author} {\bibinfo {author} {\bibfnamefont {W.~E.}\ \bibnamefont
  {Liu}}, \bibinfo {author} {\bibfnamefont {J.}~\bibnamefont {Levinsen}}, \
  and\ \bibinfo {author} {\bibfnamefont {M.~M.}\ \bibnamefont {Parish}},\
  }\bibfield  {title} {\bibinfo {title} {\emph {{Variational Approach for
  Impurity Dynamics at Finite Temperature}}},\ }\href {\doibase
  10.1103/PhysRevLett.122.205301} {\bibfield  {journal} {\bibinfo  {journal}
  {Phys. Rev. Lett.}\ }\textbf {\bibinfo {volume} {122}},\ \bibinfo {pages}
  {205301} (\bibinfo {year} {2019})}\BibitemShut {NoStop}%
\bibitem [{\citenamefont {Zhang}\ \emph {et~al.}(2015)\citenamefont {Zhang},
  \citenamefont {Cheng}, \citenamefont {Zhai},\ and\ \citenamefont
  {Zhang}}]{Zhang2015}%
  \BibitemOpen
  \bibfield  {author} {\bibinfo {author} {\bibfnamefont {R.}~\bibnamefont
  {Zhang}}, \bibinfo {author} {\bibfnamefont {Y.}~\bibnamefont {Cheng}},
  \bibinfo {author} {\bibfnamefont {H.}~\bibnamefont {Zhai}}, \ and\ \bibinfo
  {author} {\bibfnamefont {P.}~\bibnamefont {Zhang}},\ }\bibfield  {title}
  {\bibinfo {title} {\emph {Orbital Feshbach Resonance in Alkali-Earth
  Atoms}},\ }\href {\doibase 10.1103/PhysRevLett.115.135301} {\bibfield
  {journal} {\bibinfo  {journal} {Phys. Rev. Lett.}\ }\textbf {\bibinfo
  {volume} {115}},\ \bibinfo {pages} {135301} (\bibinfo {year}
  {2015})}\BibitemShut {NoStop}%
\bibitem [{sup()}]{supmat}%
  \BibitemOpen
  \href@noop {} {}\bibinfo {note} {See the supplemental material for details of
  the model and scattering parameters, the variational approach, the
  perturbative analysis, and the relationship between Rabi oscillations and the
  quasiparticle width. This includes a table of experimental parameters used
  for the TBM simulations, as well as
  Refs.~\cite{Xu2016,Kirk2017,Gurarie2007,Petrov2001,Bloch2008mbp}.}\BibitemShut
  {Stop}%
\bibitem [{\citenamefont {Liu}\ \emph {et~al.}(2020{\natexlab{a}})\citenamefont
  {Liu}, \citenamefont {Shi}, \citenamefont {Levinsen},\ and\ \citenamefont
  {Parish}}]{Liu2020short}%
  \BibitemOpen
  \bibfield  {author} {\bibinfo {author} {\bibfnamefont {W.~E.}\ \bibnamefont
  {Liu}}, \bibinfo {author} {\bibfnamefont {Z.-Y.}\ \bibnamefont {Shi}},
  \bibinfo {author} {\bibfnamefont {J.}~\bibnamefont {Levinsen}}, \ and\
  \bibinfo {author} {\bibfnamefont {M.~M.}\ \bibnamefont {Parish}},\ }\bibfield
   {title} {\bibinfo {title} {\emph {Radio-Frequency Response and Contact of
  Impurities in a Quantum Gas}},\ }\href {\doibase
  10.1103/PhysRevLett.125.065301} {\bibfield  {journal} {\bibinfo  {journal}
  {Phys. Rev. Lett.}\ }\textbf {\bibinfo {volume} {125}},\ \bibinfo {pages}
  {065301} (\bibinfo {year} {2020}{\natexlab{a}})}\BibitemShut {NoStop}%
\bibitem [{\citenamefont {Liu}\ \emph {et~al.}(2020{\natexlab{b}})\citenamefont
  {Liu}, \citenamefont {Shi}, \citenamefont {Parish},\ and\ \citenamefont
  {Levinsen}}]{Liu2020long}%
  \BibitemOpen
  \bibfield  {author} {\bibinfo {author} {\bibfnamefont {W.~E.}\ \bibnamefont
  {Liu}}, \bibinfo {author} {\bibfnamefont {Z.-Y.}\ \bibnamefont {Shi}},
  \bibinfo {author} {\bibfnamefont {M.~M.}\ \bibnamefont {Parish}}, \ and\
  \bibinfo {author} {\bibfnamefont {J.}~\bibnamefont {Levinsen}},\ }\bibfield
  {title} {\bibinfo {title} {\emph {Theory of radio-frequency spectroscopy of
  impurities in quantum gases}},\ }\href {\doibase 10.1103/PhysRevA.102.023304}
  {\bibfield  {journal} {\bibinfo  {journal} {Phys. Rev. A}\ }\textbf {\bibinfo
  {volume} {102}},\ \bibinfo {pages} {023304} (\bibinfo {year}
  {2020}{\natexlab{b}})}\BibitemShut {NoStop}%
\bibitem [{\citenamefont {Levinsen}\ and\ \citenamefont
  {Parish}(2013)}]{Levinsen2013}%
  \BibitemOpen
  \bibfield  {author} {\bibinfo {author} {\bibfnamefont {J.}~\bibnamefont
  {Levinsen}}\ and\ \bibinfo {author} {\bibfnamefont {M.~M.}\ \bibnamefont
  {Parish}},\ }\bibfield  {title} {\bibinfo {title} {\emph {Bound States in a
  Quasi-Two-Dimensional Fermi Gas}},\ }\href {\doibase
  10.1103/PhysRevLett.110.055304} {\bibfield  {journal} {\bibinfo  {journal}
  {Phys. Rev. Lett.}\ }\textbf {\bibinfo {volume} {110}},\ \bibinfo {pages}
  {055304} (\bibinfo {year} {2013})}\BibitemShut {NoStop}%
\bibitem [{\citenamefont {H\"ofer}\ \emph {et~al.}(2015)\citenamefont
  {H\"ofer}, \citenamefont {Riegger}, \citenamefont {Scazza}, \citenamefont
  {Hofrichter}, \citenamefont {Fernandes}, \citenamefont {Parish},
  \citenamefont {Levinsen}, \citenamefont {Bloch},\ and\ \citenamefont
  {F\"olling}}]{Hofer2015}%
  \BibitemOpen
  \bibfield  {author} {\bibinfo {author} {\bibfnamefont {M.}~\bibnamefont
  {H\"ofer}}, \bibinfo {author} {\bibfnamefont {L.}~\bibnamefont {Riegger}},
  \bibinfo {author} {\bibfnamefont {F.}~\bibnamefont {Scazza}}, \bibinfo
  {author} {\bibfnamefont {C.}~\bibnamefont {Hofrichter}}, \bibinfo {author}
  {\bibfnamefont {D.~R.}\ \bibnamefont {Fernandes}}, \bibinfo {author}
  {\bibfnamefont {M.~M.}\ \bibnamefont {Parish}}, \bibinfo {author}
  {\bibfnamefont {J.}~\bibnamefont {Levinsen}}, \bibinfo {author}
  {\bibfnamefont {I.}~\bibnamefont {Bloch}}, \ and\ \bibinfo {author}
  {\bibfnamefont {S.}~\bibnamefont {F\"olling}},\ }\bibfield  {title} {\bibinfo
  {title} {\emph {Observation of an Orbital Interaction-Induced Feshbach
  Resonance in $^{173}\mathrm{Yb}$}},\ }\href {\doibase
  10.1103/PhysRevLett.115.265302} {\bibfield  {journal} {\bibinfo  {journal}
  {Phys. Rev. Lett.}\ }\textbf {\bibinfo {volume} {115}},\ \bibinfo {pages}
  {265302} (\bibinfo {year} {2015})}\BibitemShut {NoStop}%
\bibitem [{\citenamefont {Pagano}\ \emph {et~al.}(2015)\citenamefont {Pagano},
  \citenamefont {Mancini}, \citenamefont {Cappellini}, \citenamefont {Livi},
  \citenamefont {Sias}, \citenamefont {Catani}, \citenamefont {Inguscio},\ and\
  \citenamefont {Fallani}}]{Pagano2015}%
  \BibitemOpen
  \bibfield  {author} {\bibinfo {author} {\bibfnamefont {G.}~\bibnamefont
  {Pagano}}, \bibinfo {author} {\bibfnamefont {M.}~\bibnamefont {Mancini}},
  \bibinfo {author} {\bibfnamefont {G.}~\bibnamefont {Cappellini}}, \bibinfo
  {author} {\bibfnamefont {L.}~\bibnamefont {Livi}}, \bibinfo {author}
  {\bibfnamefont {C.}~\bibnamefont {Sias}}, \bibinfo {author} {\bibfnamefont
  {J.}~\bibnamefont {Catani}}, \bibinfo {author} {\bibfnamefont
  {M.}~\bibnamefont {Inguscio}}, \ and\ \bibinfo {author} {\bibfnamefont
  {L.}~\bibnamefont {Fallani}},\ }\bibfield  {title} {\bibinfo {title} {\emph
  {Strongly Interacting Gas of Two-Electron Fermions at an Orbital Feshbach
  Resonance}},\ }\href {\doibase 10.1103/PhysRevLett.115.265301} {\bibfield
  {journal} {\bibinfo  {journal} {Phys. Rev. Lett.}\ }\textbf {\bibinfo
  {volume} {115}},\ \bibinfo {pages} {265301} (\bibinfo {year}
  {2015})}\BibitemShut {NoStop}%
\bibitem [{\citenamefont {Bishop}(1973)}]{Bishop1973}%
  \BibitemOpen
  \bibfield  {author} {\bibinfo {author} {\bibfnamefont {R.}~\bibnamefont
  {Bishop}},\ }\bibfield  {title} {\bibinfo {title} {\emph {{On the ground
  state of an impurity in a dilute fermi gas}}},\ }\href {\doibase
  10.1016/0003-4916(73)90265-0} {\bibfield  {journal} {\bibinfo  {journal}
  {Ann. Phys. (N.~Y.)}\ }\textbf {\bibinfo {volume} {78}},\ \bibinfo {pages}
  {391} (\bibinfo {year} {1973})}\BibitemShut {NoStop}%
\bibitem [{\citenamefont {Petrov}(2003)}]{Petrov2003}%
  \BibitemOpen
  \bibfield  {author} {\bibinfo {author} {\bibfnamefont {D.~S.}\ \bibnamefont
  {Petrov}},\ }\bibfield  {title} {\bibinfo {title} {\emph {Three-body problem
  in Fermi gases with short-range interparticle interaction}},\ }\href
  {\doibase 10.1103/PhysRevA.67.010703} {\bibfield  {journal} {\bibinfo
  {journal} {Phys. Rev. A}\ }\textbf {\bibinfo {volume} {67}},\ \bibinfo
  {pages} {010703} (\bibinfo {year} {2003})}\BibitemShut {NoStop}%
\bibitem [{\citenamefont {Parish}\ and\ \citenamefont
  {Levinsen}(2016)}]{Parish2016}%
  \BibitemOpen
  \bibfield  {author} {\bibinfo {author} {\bibfnamefont {M.~M.}\ \bibnamefont
  {Parish}}\ and\ \bibinfo {author} {\bibfnamefont {J.}~\bibnamefont
  {Levinsen}},\ }\bibfield  {title} {\bibinfo {title} {\emph {{Quantum dynamics
  of impurities coupled to a Fermi sea}}},\ }\href {\doibase
  10.1103/PhysRevB.94.184303} {\bibfield  {journal} {\bibinfo  {journal} {Phys.
  Rev. B}\ }\textbf {\bibinfo {volume} {94}},\ \bibinfo {pages} {184303}
  (\bibinfo {year} {2016})}\BibitemShut {NoStop}%
\bibitem [{\citenamefont {Goulko}\ \emph {et~al.}(2017)\citenamefont {Goulko},
  \citenamefont {Mishchenko}, \citenamefont {Pollet}, \citenamefont
  {Prokof'ev},\ and\ \citenamefont {Svistunov}}]{Goulko2017}%
  \BibitemOpen
  \bibfield  {author} {\bibinfo {author} {\bibfnamefont {O.}~\bibnamefont
  {Goulko}}, \bibinfo {author} {\bibfnamefont {A.~S.}\ \bibnamefont
  {Mishchenko}}, \bibinfo {author} {\bibfnamefont {L.}~\bibnamefont {Pollet}},
  \bibinfo {author} {\bibfnamefont {N.}~\bibnamefont {Prokof'ev}}, \ and\
  \bibinfo {author} {\bibfnamefont {B.}~\bibnamefont {Svistunov}},\ }\bibfield
  {title} {\bibinfo {title} {\emph {Numerical analytic continuation: Answers to
  well-posed questions}},\ }\href {\doibase 10.1103/PhysRevB.95.014102}
  {\bibfield  {journal} {\bibinfo  {journal} {Phys. Rev. B}\ }\textbf {\bibinfo
  {volume} {95}},\ \bibinfo {pages} {014102} (\bibinfo {year}
  {2017})}\BibitemShut {NoStop}%
\bibitem [{\citenamefont {Xu}\ \emph {et~al.}(2016)\citenamefont {Xu},
  \citenamefont {Zhang}, \citenamefont {Cheng}, \citenamefont {Zhang},
  \citenamefont {Qi},\ and\ \citenamefont {Zhai}}]{Xu2016}%
  \BibitemOpen
  \bibfield  {author} {\bibinfo {author} {\bibfnamefont {J.}~\bibnamefont
  {Xu}}, \bibinfo {author} {\bibfnamefont {R.}~\bibnamefont {Zhang}}, \bibinfo
  {author} {\bibfnamefont {Y.}~\bibnamefont {Cheng}}, \bibinfo {author}
  {\bibfnamefont {P.}~\bibnamefont {Zhang}}, \bibinfo {author} {\bibfnamefont
  {R.}~\bibnamefont {Qi}}, \ and\ \bibinfo {author} {\bibfnamefont
  {H.}~\bibnamefont {Zhai}},\ }\bibfield  {title} {\bibinfo {title} {\emph
  {Reaching a Fermi-superfluid state near an orbital Feshbach resonance}},\
  }\href {\doibase 10.1103/PhysRevA.94.033609} {\bibfield  {journal} {\bibinfo
  {journal} {Phys. Rev. A}\ }\textbf {\bibinfo {volume} {94}},\ \bibinfo
  {pages} {033609} (\bibinfo {year} {2016})}\BibitemShut {NoStop}%
\bibitem [{\citenamefont {Kirk}\ and\ \citenamefont {Parish}(2017)}]{Kirk2017}%
  \BibitemOpen
  \bibfield  {author} {\bibinfo {author} {\bibfnamefont {T.}~\bibnamefont
  {Kirk}}\ and\ \bibinfo {author} {\bibfnamefont {M.~M.}\ \bibnamefont
  {Parish}},\ }\bibfield  {title} {\bibinfo {title} {\emph {Three-body
  correlations in a two-dimensional SU(3) Fermi gas}},\ }\href {\doibase
  10.1103/PhysRevA.96.053614} {\bibfield  {journal} {\bibinfo  {journal} {Phys.
  Rev. A}\ }\textbf {\bibinfo {volume} {96}},\ \bibinfo {pages} {053614}
  (\bibinfo {year} {2017})}\BibitemShut {NoStop}%
\bibitem [{\citenamefont {Gurarie}\ and\ \citenamefont
  {Radzihovsky}(2007)}]{Gurarie2007}%
  \BibitemOpen
  \bibfield  {author} {\bibinfo {author} {\bibfnamefont {V.}~\bibnamefont
  {Gurarie}}\ and\ \bibinfo {author} {\bibfnamefont {L.}~\bibnamefont
  {Radzihovsky}},\ }\bibfield  {title} {\bibinfo {title} {\emph {Resonantly
  paired fermionic superfluids}},\ }\href {\doibase
  https://doi.org/10.1016/j.aop.2006.10.009} {\bibfield  {journal} {\bibinfo
  {journal} {Ann. Phys. (N.~Y.)}\ }\textbf {\bibinfo {volume} {322}},\ \bibinfo
  {pages} {2 } (\bibinfo {year} {2007})}\BibitemShut {NoStop}%
\bibitem [{\citenamefont {Petrov}\ and\ \citenamefont
  {Shlyapnikov}(2001)}]{Petrov2001}%
  \BibitemOpen
  \bibfield  {author} {\bibinfo {author} {\bibfnamefont {D.~S.}\ \bibnamefont
  {Petrov}}\ and\ \bibinfo {author} {\bibfnamefont {G.~V.}\ \bibnamefont
  {Shlyapnikov}},\ }\bibfield  {title} {\bibinfo {title} {\emph {Interatomic
  collisions in a tightly confined Bose gas}},\ }\href {\doibase
  10.1103/PhysRevA.64.012706} {\bibfield  {journal} {\bibinfo  {journal} {Phys.
  Rev. A}\ }\textbf {\bibinfo {volume} {64}},\ \bibinfo {pages} {012706}
  (\bibinfo {year} {2001})}\BibitemShut {NoStop}%
\bibitem [{\citenamefont {Bloch}\ \emph {et~al.}(2008)\citenamefont {Bloch},
  \citenamefont {Dalibard},\ and\ \citenamefont {Zwerger}}]{Bloch2008mbp}%
  \BibitemOpen
  \bibfield  {author} {\bibinfo {author} {\bibfnamefont {I.}~\bibnamefont
  {Bloch}}, \bibinfo {author} {\bibfnamefont {J.}~\bibnamefont {Dalibard}}, \
  and\ \bibinfo {author} {\bibfnamefont {W.}~\bibnamefont {Zwerger}},\
  }\bibfield  {title} {\bibinfo {title} {\emph {Many-body physics with
  ultracold gases}},\ }\href {\doibase 10.1103/RevModPhys.80.885} {\bibfield
  {journal} {\bibinfo  {journal} {Rev. Mod. Phys.}\ }\textbf {\bibinfo {volume}
  {80}},\ \bibinfo {pages} {885} (\bibinfo {year} {2008})}\BibitemShut
  {NoStop}%
\end{thebibliography}%



\renewcommand{\theequation}{S\arabic{equation}}
\renewcommand{\thefigure}{S\arabic{figure}}
\renewcommand{\thetable}{S\arabic{table}}

\onecolumngrid


\setcounter{equation}{0}
\setcounter{figure}{0}
\setcounter{table}{0}

\clearpage

\section*{SUPPLEMENTAL MATERIAL:\\ ``Quasiparticle lifetime of the repulsive Fermi polaron''}
\setcounter{page}{1}
\begin{center}
Haydn S. Adlong,$^1$
Weizhe Edward Liu,$^{1,2}$
Francesco Scazza,$^3$
Matteo Zaccanti,$^3$
Nelson Darkwah Oppong,$^{4,5,6}$
Simon~F\"olling,$^{4,5,6}$
Meera M.~Parish,$^{1,2}$ and
Jesper~Levinsen$^{1,2}$, \\
\emph{\small $^1$School of Physics and Astronomy, Monash University, Victoria 3800, Australia}\\
\emph{\small $^2$ARC Centre of Excellence in Future Low-Energy Electronics Technologies, Monash University, Victoria 3800, Australia}\\
\emph{\small $^3$Istituto Nazionale di Ottica del Consiglio Nazionale delle Ricerche (CNR-INO) and European Laboratory for Nonlinear Spectroscopy (LENS), 50019 Sesto Fiorentino, Italy}\\
\emph{\small $^4$Ludwig-Maximilians-Universit{\"a}t, Schellingstra{\ss}e 4, 80799 M{\"u}nchen, Germany}\\
\emph{\small $^5$Max-Planck-Institut f{\"u}r Quantenoptik, Hans-Kopfermann-Stra{\ss}e 1, 85748 Garching, Germany}\\
\emph{\small $^6$Munich Center for Quantum Science and Technology (MCQST), Schellingstra{\ss}e 4, 80799 M{\"u}nchen, Germany}

\end{center}

\section{Model and scattering parameters}

\subsection{Model}

In modelling the dynamics of impurities coupled to a Fermi sea, we use the following effective Hamiltonian
\begin{align} \label{eq:effectiveHam}
\hat H=\hat H_0+\hat H_\Omega+\hat H_\up+\hat H_\down,
\end{align}
where
\begin{subequations}
\begin{align}
    \hat H_0&=\sum_\k(\ek-\mu) \hat f_\k^\dag\hat f_\k, \label{eq:supHam1}\\
   \hat H_\Omega&=\frac{\Omega_0}{2} \sum_{\k} \left( 
   \chd_{\k \down} \ch_{\k \up} + 
   \chd_{\k \up} \ch_{\k \down} \right)+\Delta\omega\,\hat n_{\down}, \label{eq:supHam2}\\
    \hat{H}_{\sigma} &= \sum_{\k} \left[\epsilon_{\k} \hat
                       c^\dag_{\k\sigma} \hat c_{\k\sigma} +
                       ( \ek/2 + \nu_{\sigma}) \hat d^\dag_{\k\sigma} \hat d_{\k\sigma}\right]                  
+g_{\sigma}\sum_{\k, \q} \left( \hat d^\dag_{\q\sigma}\hat c_{\q/2 - \k,\sigma}  \hat f_{\q/2 + \k}
                       + \hat f^\dag_{\q/2 + \k} \hat c^\dag_{\q/2
                       -\k, \sigma}\hat d_{\q\sigma} \right).\label{eq:supHam3}
\end{align}
\end{subequations}
The meaning of the various symbols is discussed in the main text.

\subsection{Relation between Hamiltonian operators and experimental atomic states}

In the 2D $^{173}$Yb experiment of Ref.~\cite{Oppong2019}, the states of the atoms are defined by the electronic state, with ground state $^1S_0$ and long-lived excited state $^3P_0$ (i.e., the ``clock'' state), as well as the nuclear-spin state with $m_F \in \{-5/2, -3/2, \ldots, +5/2\}$.
The majority $\hat f_\k^\dag$ atoms exist in the $m_F=+5/2$ ground state, which acts as a bath for the weakly interacting $\chd_{\k \downarrow}$ and resonantly interacting $\chd_{\k \uparrow}$ impurities in the $m_F=-3/2$ ground state and $m_F=-5/2$ `clock' state, respectively.
Interactions between the $\chd_{\k \uparrow}$ and $\hat f_\k^\dag$ atoms are tunable through an orbital Feshbach resonance~\cite{Zhang2015} as has been demonstrated experimentally~\cite{Hofer2015,Pagano2015}.

On the other hand, the 3D experiment of Ref.~\cite{Scazza2017} involves $^6$Li atoms in the three lowest Zeeman levels. The majority $\hat f_\k^\dag$ atoms exist in the lowest Zeeman level, while the weakly interacting $\chd_{\k \downarrow}$ and resonantly interacting $\chd_{\k \uparrow}$ impurities occupy the second-lowest and third-lowest Zeeman levels respectively. Owing to two off-centered broad Feshbach resonances, the scattering between the  $\chd_{\k \uparrow}$ and $\hat f_\k^\dag$ atoms can be resonantly enhanced while only moderately increasing the comparatively weak interactions between the $\chd_{\k \downarrow}$ and $\hat f_\k^\dag$ atoms~\cite{Scazza2017}.

\subsection{Scattering parameters}

Within the model in Eq.~\eqref{eq:effectiveHam}, fermions of the same spin do not interact and the two spin states of the impurity are only coupled through the light-field. Ignoring for the moment the light-field (i.e., setting $\Omega_0=0$), we calculate the vacuum spin-dependent impurity-fermion $T$ matrix, which characterizes the interactions between the majority fermions and a particular spin state of the impurity. %
This yields
\begin{align}
    T_{\sigma}(E)=\left[\frac{E - \nu_\sigma}{g^2_{\sigma}}-\sum_{\k}^\Lambda \frac{1}{E-2 \epsilon_{\k}}\right]^{-1}, %
    \label{eq:Tmatrix}
\end{align}
where $E$ is the collision energy and $\Lambda$ is the ultraviolet cutoff.

To proceed, we compare the scattering $T$ matrix with the 2D and 3D scattering amplitudes using $f_{2{\rm D}\sigma}(k)=mT_\sigma(E)$ and $f_{3{\rm D}\sigma}(k)=-\frac{m}{4\pi}T_\sigma(E)$, with $E=k^2/m$. The scattering amplitudes at low energy are known to take the forms
\begin{subequations}
\begin{align} 
     f_{\text{2D}\sigma}(k) &\simeq   \frac{4 \pi}{ - \ln (k^2 a^2_{\text{2D}\sigma}) + R^2_{\text{2D}\sigma}k^2+i \pi},
     \label{eq:f2D}\\
     f_{\text{3D}\sigma}(k) &\simeq -\frac{1}{ a_{\text{3D}\sigma}^{-1} + R_{\text{3D}\sigma} k^2 + ik }.
\end{align}
\end{subequations}
By comparing Eq.~\eqref{eq:Tmatrix} with the low-energy scattering amplitudes, we obtain the renormalized scattering parameters. In 2D, this results in
\begin{align}
\frac{\nu_\sigma+E_{2\sigma}}{g^2_{\sigma}} = \sum_{\k}^\Lambda \frac{1}{E_{2\sigma}+2 \epsilon_{\k}},  \qquad R_{\text{2D}\sigma}^2  = \frac{4\pi}{m^2 g^2_\sigma},
\end{align}
where $a_{{\rm 2D}\sigma}$ is the scattering length, $R_{2{\rm D}\sigma}$ is the effective range~\cite{Kirk2017}, and $E_{2\sigma}$ is the binding energy of the two-body bound state that exists for all interactions in 2D:
\begin{align}
    E_{2 \sigma} = \frac{1}{m R^2_{\text{2D}\sigma}} W\left( \frac{R^2_{\text{2D}\sigma}}{a^2_{\text{2D}\sigma}} \right),
\end{align}
with $W$ the Lambert $W$ function.
In 3D, we have
\begin{align}
    \frac{m}{4 \pi a_{\text{3D}\sigma}} = - \frac{\nu_\sigma}{g^2_\sigma} + \sum_{\k}^\Lambda \frac{1}{ 2 \epsilon_{\k}}, \qquad R_{\text{3D} \sigma} = \frac{4 \pi}{m^2 g_\sigma^2},
    \label{eq:renorm3D}
\end{align}
where $a_{{\rm 3D}\sigma}$ is the scattering length and $R_{3{\rm D}\sigma}$ is a range parameter~\cite{Gurarie2007}. In this case, we only have a bound state when $a_{{\rm 3D}\sigma}>0$, with corresponding binding energy
\begin{align}
    E_{3\sigma}=\frac{\left[\sqrt{1+4R_{3{\rm D}\sigma}/a_{3{\rm D}\sigma}}-1\right]^2}{4m{R_{3{\rm D}\sigma}^{2}}}.
\end{align}
Through this procedure, we have related the bare interaction parameters $g$, $\Lambda$, and $\nu$ to the physical parameters, the scattering length $a$ and effective range $R$, which characterises the relevant Feshbach resonance. For the broad Feshbach resonances in ${}^6$Li we use $R_{\text{3D}\sigma}=0$ for both spin states. For the orbital Feshbach resonance in quasi-2D, the description of the effective range is somewhat more complicated, as discussed in the following.

\subsection{Effective model for scattering at an orbital Feshbach resonance in the presence of confinement}

We now discuss how the scattering parameters, $a_{\text{2D}\sigma}$ and $R_{\text{2D}\sigma}$, of the $^{173}$Yb experiment are determined. The orbital Feshbach resonance is known to lead to a strongly energy dependent scattering~\cite{Zhang2015}, which is well approximated by the introduction of a 3D effective range~\cite{Xu2016}. Furthermore, quasi-2D confinement generally leads to a non-trivial energy dependence of the effective 2D scattering amplitude even for a broad Feshbach resonance~\cite{Petrov2001}; for strong confinement, this energy dependence can be modelled by the introduction of a 2D effective range~\cite{Levinsen2013}. Thus, we use the effective range in our model to provide the simplest possible description of both the orbital Feshbach resonance and the confinement. This greatly simplifies the numerical simulations of Rabi oscillations, since it drastically reduces the possible degrees of freedom in the problem.

We now describe the possible scattering channels in ${}^{173}$Yb atoms close to an orbital Feshbach resonance, following the analysis in Ref.~\cite{Zhang2015}. We focus on the two electronic orbitals described above (denoted here by $\ket{g}$ and $\ket{e}$) as well as two particular nuclear spin states (denoted here by $\ket{\Downarrow}$ and $\ket{\Uparrow}$). These states form the open channel $\ket{o} \equiv \ket{g \Uparrow, e \Downarrow}$ and the closed channel $\ket{c} \equiv \ket{e \Uparrow, g \Downarrow}$, which are detuned by $\delta = \Delta \mu B$, where $\Delta \mu = h\times 554\,\mathrm{Hz}/\mathrm{G}$~\cite{Oppong2019} is the differential Zeeman shift and $B$ is the magnetic field strength in Gauss.
The interactions in this system are not diagonal in the open- and closed-channel basis, but instead proceed via the triplet $(+)$ and singlet $(-)$ channels, with $\ket{\pm} \equiv \frac{1}{\sqrt{2}}(\ket{o} \pm \ket{c})$. Associated with the singlet and triplet interactions are the singlet and triplet scattering lengths $a_\pm$~\cite{Zhang2015} and effective ranges $r_\pm$~\cite{Hofer2015}. 

\begin{figure}[t]
    \centering
    \includegraphics[width=0.7\linewidth]{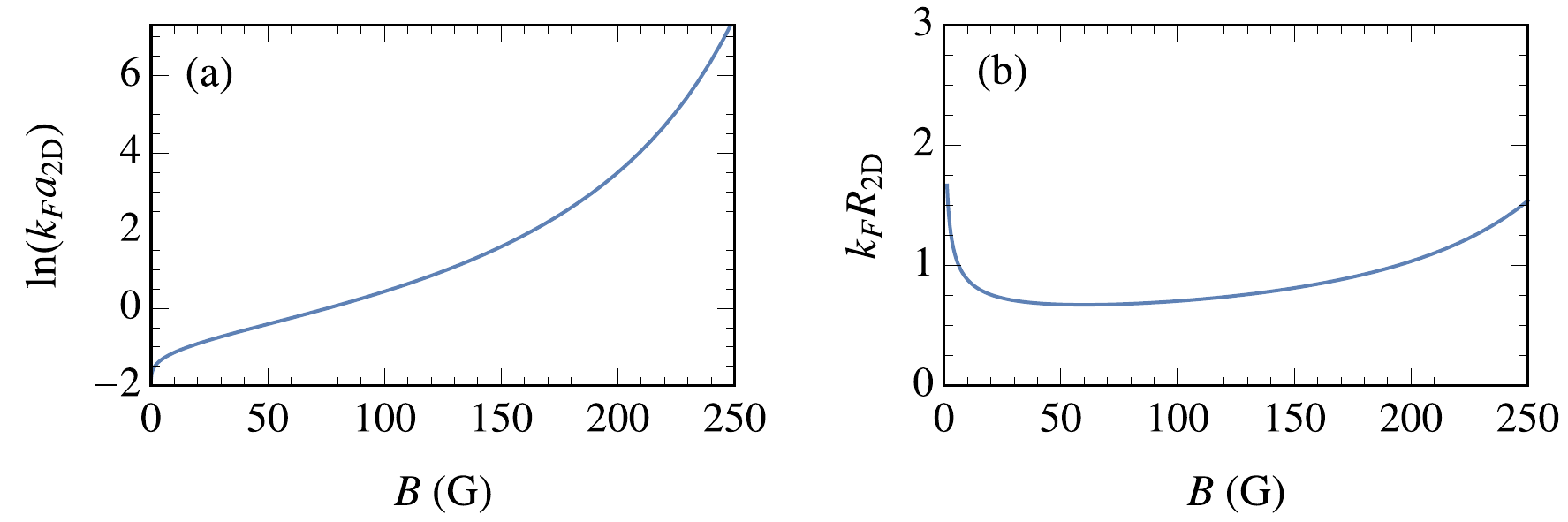}
    \caption{The open channel 2D scattering length (a) and 2D effective range (b) as a function of magnetic field strength (in Gauss). These parameters are used to create an effective 2D two-channel model description of scattering at an orbital Feshbach resonance in the presence of confinement. To match Ref.~\cite{Oppong2019} we use $\omega_z = 2\pi \times 37.1\,\mathrm{kHz}$.}
    \label{fig:a2DandR2D}
\end{figure}

In the experiment~\cite{Oppong2019}, the ${}^{173}$Yb atoms were confined to move in a 2D plane with a harmonic potential $V(z) = \frac{1}{2} m \omega_z^2 z^2$ acting in the transverse direction. Reference~\cite{Oppong2019} used this to extend the theoretical analysis of quasi-2D scattering in Refs.~\cite{Petrov2001,Bloch2008mbp} to derive an effective scattering amplitude of the form in Eq.~\eqref{eq:f2D}. For the weakly interacting $\down$ state, it was found that $\ln (k_F a_{\text{2D}\downarrow}) \simeq -4.9 (1)$~\cite{Oppong2019}. In this case, the dependence on $R_{\text{2D} \downarrow}$ is strongly suppressed, and we simply take $R_{\text{2D} \downarrow}=0$. For the case of strong interactions, the open channel scattering amplitude was found to be given by~\cite{Oppong2019}
\begin{align}
    f_{\text{q2D}}(E) &=2 \sqrt{2 \pi } \frac{ l_z \left(\Tilde{a}^{-1}_- -  \frac{m r_- E}{2} \right)+ l_z \left(\Tilde{a}^{-1}_+ -  \frac{m  r_+ E}{2} \right)-2 \mathcal{F}\left(\frac{-E + \delta}{\omega_z}\right)}{ \left\{\splitfrac{ -\mathcal{F}\left(\frac{-E + \delta}{\omega_z}\right) \left[l_z (\Tilde{a}^{-1}_- -  \frac{m r_- E }{2})+l_z (\Tilde{a}^{-1}_+ -  \frac{m r_+ E }{2})-2
   \mathcal{F}\left(-\frac{E}{\omega_z}\right) \right]}{-\mathcal{F}\left(-\frac{E}{\omega_z}\right) \left[l_z (\Tilde{a}^{-1}_- -  \frac{m r_- E }{2})+l_z (\Tilde{a}^{-1}_+ -  \frac{m r_+ E}{2}) \right] +2 l_z^2 (\Tilde{a}^{-1}_- - \frac{m  r_- E}{2}) (\Tilde{a}^{-1}_+ -  \frac{ m r_+ E }{2}) } \right\}}, \label{eq:fopen}
\end{align}
where $l_z = 1/\sqrt{m \omega_z}$ and $\Tilde{a}_\pm^{-1} = a^{-1}_\pm - \frac{r_\pm}{4 l_z^2}  \left( 1- \delta/\omega_z \right)$. Here, $\mathcal{F}$ is a transcendental function defined by~\cite{Bloch2008mbp}
\begin{align}
    \mathcal{F}(x) = \int_0^\infty \frac{du}{\sqrt{4 \pi u^3}} \left[ 1 - \frac{e^{-xu}}{\sqrt{(1-e^{-2u})/2u}} \right],
\end{align}
and the energy is measured with respect to the quasi-2D zero point energy.

To extract an effective low-energy open-channel 2D scattering length and effective range for the strongly interacting spin-$\up$ impurity case, we perform a low energy expansion of  Eq.~\eqref{eq:fopen} and compare it with the standard form of the low-energy scattering amplitude, Eq.~\eqref{eq:f2D}. In the following, we use the notation $a_{\text{2D}}\equiv a_{\text{2D}\uparrow}$ and $R_{\text{2D}}\equiv R_{\text{2D}\uparrow}$, as in the main text. Assuming that $\delta \gg |E|$ (i.e., that we are not close to $B=0$) we find the 2D scattering length \cite{Oppong2019}
\begin{align}
    a_{\text{2D}} = l_z \sqrt{\frac{\pi}{D}} \exp[-\sqrt{2 \pi} \frac{ l_z^2 ( \Tilde{a}_{-} \Tilde{a}_{+})^{-1}- \frac{1}{2}(l_z \Tilde{a}_{-}^{-1}+ l_z \Tilde{a}_{+}^{-1}) \mathcal{F}\left(\frac{\delta}{\omega_z} \right) }{l_z \Tilde{a}_{-}^{-1}+l_z \Tilde{a}_{+}^{-1}-2 \mathcal{F} \left(\frac{\delta}{\omega_z}
   \right)} ],
\end{align}
where $D \simeq 0.905$~\cite{Petrov2001}. The 2D effective range takes the form
\begin{align}
    \left(\frac{R_{\text{2D}}}{l_z}\right)^2 &=  \ln 2 %
    -\frac{\sqrt{2 \pi } \left\{ l_z^{-1} r_+ \left[ l_z \Tilde{a}_-^{-1} - \mathcal{F}\left(\frac{\delta}{\omega_z}
   \right) \right]^2 + l_z^{-1} r_- \left[l_z \Tilde{a}_+^{-1} - \mathcal{F}\left(\frac{\delta}{\omega_z}
   \right) \right]^2  + \left[l_z\Tilde{a}_-^{-1} - l_z \Tilde{a}_+^{-1} \right]^2 \mathcal{F}'\left( \frac{\delta}{\omega_z} \right)\right\}}{ \left(l_z \Tilde{a}_-^{-1}+ l_z\Tilde{a}_+^{-1} -2 \mathcal{F}\left( \frac{\delta}{\omega_z} \right) \right)^2},
\end{align}
where $\mathcal{F}'$ is the first derivative of $\mathcal{F}$. 

Figure~\ref{fig:a2DandR2D} shows the extracted scattering parameters using the experimentally determined values for all parameters~\cite{Oppong2019} (see also Ref.~\cite{Hofer2015}). We see that indeed the change of the magnetic field provides a means to tune the interactions via $a_{{\rm 2D}}$. On the other hand, the effective range is not strongly dependent on magnetic field --- indeed, the effective range is approximately constant ($k_F R_{\text{2D}} \sim 1$) within the magnetic field strengths of interest to the present work.

\begin{figure}
    \centering
    \includegraphics[width=0.35\linewidth]{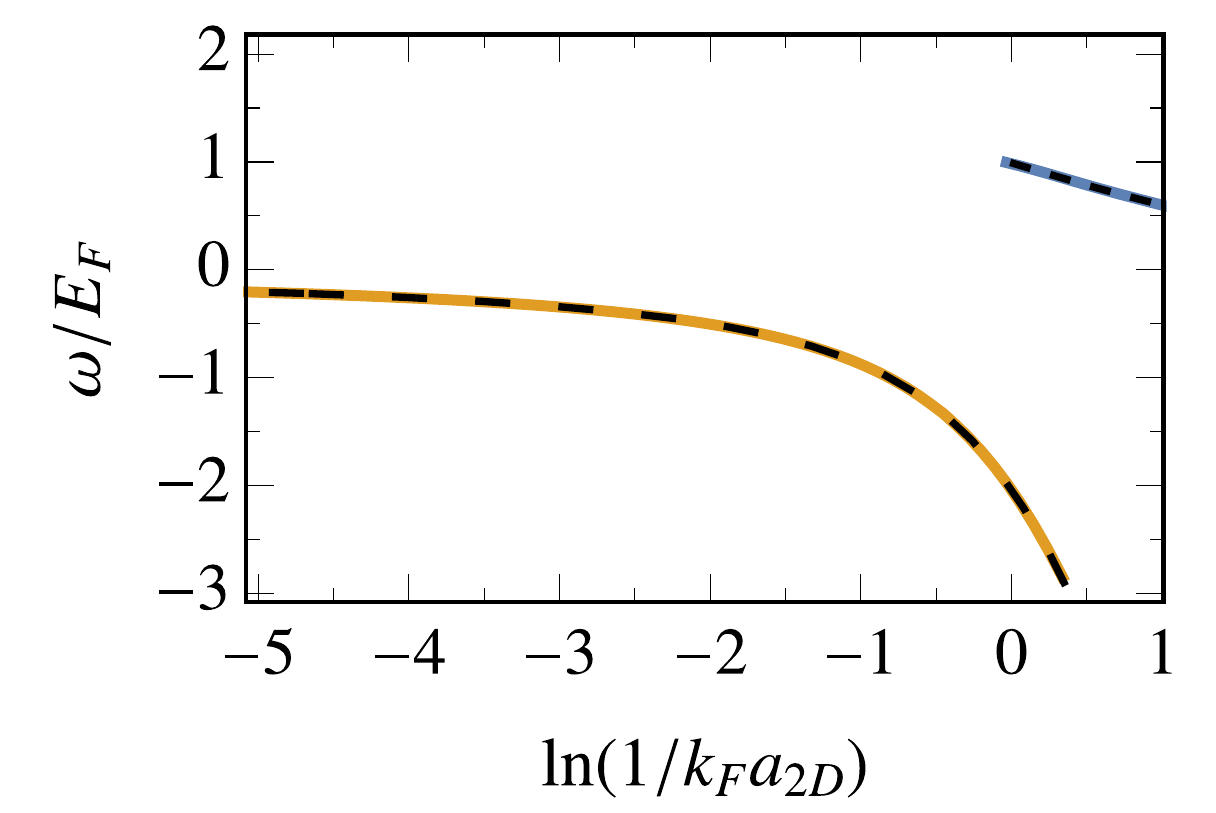}
    \caption{\jfl{
    The attractive (orange) and repulsive (blue) Fermi polaron energy as calculated through our effective two-channel model description of an orbital Feshbach resonance in the presence of confinement. The energies are highly consistent with the full theory calculation of the polaron energies (black) from Ref. \cite{Oppong2019}. The curves are calculated at temperature $T/T_F=0.16$.}}
    \label{fig:fulltheoryvs2channel}
\end{figure}

\jfl{
In general, we find that the 2D effective range is able to highly accurately capture the physics of the orbital Feshbach resonance and the confinement in the domain of interaction strengths used in Ref.~\cite{Oppong2019}. This is shown in Fig.~\ref{fig:fulltheoryvs2channel}, where we show the near perfect agreement between the full theory model \cite{Oppong2019} and our effective model in calculating the energy of the Fermi polaron.}

\section{Finite temperature variational approach for impurity dynamics}
In order to model the dynamics of the Rabi oscillations and the impurity spectral response, we use the finite-temperature Truncated Basis Method (TBM) developed in Refs.~\cite{Parish2016,Liu2019}. The basic details of this method are discussed in the main text. Our ansatz for the impurity operator is
\begin{align} \label{eq:OpApprox}
    \hat{c}(t)= &
\sum_{\sigma}\Bigg[\alpha_{0}^\sigma(t)\hat
  c_{\0\sigma}+\sum_\k\alpha_{\k}^\sigma(t)\fhd_\k\dh_{\k\sigma}
+\sum_{\k,\q} \alpha_{\k\q}^\sigma(t) \fhd_{\q} \fh_{\k} \ch_{\q-\k \sigma}\Bigg],
\end{align}
where, for simplicity, we consider a vanishing total
momentum. As discussed in the main text, we may quantify the error incurred in the Heisenberg equation of motion by introducing the error operator $\hat\epsilon(t)\equiv i\partial_t \hat{c}(t)- \comm*{\hat{c}(t)}{\hat H}$ and the associated error quantity $\Delta(t)\equiv\Tr[\hat \rho_0 \hat\epsilon(t) \hat\epsilon^\dag(t)]$. Using the minimization condition $\pdv*{\Delta(t)}{\dot\alpha^{\sigma*}_j(t)}=0$ with respect to the variational coefficients $\{\alpha_j\}$, 
we arrive at 
\begin{subequations} \label{eq:VariationalEqs}
\begin{align}
    E  \alpha^{\uparrow}_{0} &= g_\up \sum_{\q} \alpha^{\uparrow}_{\q} \Tr[\hat \rho_0 \fhd_{\q } \fh_{\q }]  + \frac{\Omega_0}{2} \alpha^{\downarrow}_{0}\\
    (E - \varepsilon_{\q \uparrow}) \alpha^{\uparrow}_{\q} &= g_\up \alpha^{\uparrow}_{0} + g_\up \sum_{\k} \alpha^{\uparrow}_{\k \q} \Tr[\hat \rho_0 \fh_{\k}\fhd_{\k}]\\
    (E - \varepsilon_{\k \q}) \alpha^{\uparrow}_{\k \q} &= g_\up \alpha^{\uparrow}_{\q} + \frac{\Omega_0}{2} \alpha^{\downarrow}_{\k \q}\\
    (E - \Delta \omega)  \alpha^{\downarrow}_{0} &= 
    g_\down \sum_{\q} \alpha^{\downarrow}_{\q} \Tr[\hat \rho_0 \fhd_{\q } \fh_{\q }]  + \frac{\Omega_0}{2} \alpha^{\uparrow}_{0} \\
    (E - \varepsilon_{\q \downarrow} - \Delta \omega) \alpha^{\downarrow}_{\q} &= g_\down \alpha^{\downarrow}_{0}  + g_\down \sum_{\k} \alpha^{\downarrow}_{\k \q} \Tr[\hat \rho_0 \fh_{\k}\fhd_{\k}]\\
    (E - \varepsilon_{\k \q} - \Delta \omega) \alpha^{\downarrow}_{\k \q} &= g_\down \alpha^{\downarrow}_{\q} + \frac{\Omega_0}{2} \alpha^{\uparrow}_{\k \q}.
\end{align}
\end{subequations}
Here we have taken the stationary condition since our Hamiltonian is time-independent,
\begin{align}
    \alpha^\sigma_{j}(t)=\alpha^\sigma_{j}(0)e^{-iEt}\equiv\alpha^\sigma_{j}e^{-iEt},
\end{align}
and defined $\varepsilon_{\q \sigma} \equiv  \nu_{\sigma}-\epsilon_{\q}/2$ and $\varepsilon_{\k \q} \equiv \epsilon_{\q - \k} + \epsilon_{\k} - \epsilon_{\q}$.

Equation~\eqref{eq:VariationalEqs} represents a set of linear integral equations that define the time dependence of the variational coefficients $\{ \alpha_j \}$. It is quite similar to the corresponding set of equations derived in Refs.~\cite{Parish2016,Liu2019}. However, 
here we account for  temperature, initial-state interactions, and the Rabi coupling between the two impurity spin states. 
Diagonalizing Eq.~\eqref{eq:VariationalEqs} yields eigenvectors $\{ \alpha_j^{(l)} \}$ and corresponding eigenvalues $E_l$. In what follows it is useful to write the eigenvectors as the union of the spin-$\uparrow$ and spin-$\downarrow$ components, i.e., $\{ \alpha_j ^{(l)}\} = \{ \alpha^{ \uparrow (l)}_j \} \cup \{ \alpha^{ \downarrow (l)}_j \}.$

The solutions of Eq.~\eqref{eq:VariationalEqs} allow us to obtain stationary impurity operators
\begin{align}
    \hat{\phi}^{(l)} \equiv \sum_j \alpha_j^{(l)} \hat{O}_j,
\end{align}
where the impurity basis operators $\{ \hat{O}_j \}=\{\hat c_{\0\up}, \fhd_\k\dh_{\k\up},  \fhd_{\q} \fh_{\k} \ch_{\q-\k \up},\hat c_{\0\down}, \fhd_\k\dh_{\k\down},  \fhd_{\q} \fh_{\k} \ch_{\q-\k \down}\}$ are those introduced in Eq.~\eqref{eq:OpApprox}. These all satisfy $\Tr[\hat\rho_0\hat{O}_j \hat{O}^\fix{\dag}_k]=0$ when $j\neq k$, since the trace is over medium-only states. This in turn allows us to normalize the stationary solutions according to
\begin{align}
    \Tr[\hat \rho_0 \hat \phi^{(l)}\hat \phi^{(m)\dag}]=\delta_{lm}.
\end{align}
 Using this, the impurity annihilation operator in Eq.~\eqref{eq:OpApprox} can be expressed as
\begin{align}
    \hat{c}(t) = \sum_l \Tr[\hat \rho_0\ch(0)  \Hat{\phi}^{(l)\dagger}] \Hat{\phi}^{(l)} e^{-i E_l t}=\sum_l \alpha_0^{\fix{\down}(l)*} \Hat{\phi}^{(l)} e^{-i E_l t},
\end{align}
where the initial impurity operator $\ch(0) = \hat c_{\0\down}$. We take a bare impurity as the initial state even though there are initial-state interactions, since the simulations yield essentially the same result as when we use a weakly interacting polaron. We presume this is because the weakly interacting polaron forms quickly (on a time scale set by $a_{\mathrm{3D}\downarrow}^2$ in 3D) once we start the dynamics.

Finally, we discuss the character of the medium-only states appearing in Eq.~\eqref{eq:VariationalEqs}. Assuming that the interactions between the initial $\down$ impurity and the medium particles are negligible (as is the case in the experiments~\cite{Scazza2017,Oppong2019}), we may take the medium states to be thermal eigenstates at temperature $T$. Therefore, we have
\begin{align}
    \Tr[\hat \rho_0 \fhd_{\q } \fh_{\q }]\equiv \expval{\fhd_{\q } \fh_{\q }}_\beta = n_F(\eq)   
\end{align}
where $\beta$ is the inverse temperature and $n_F$ is the Fermi-Dirac distribution function:
\begin{align}
    n_F(\eq)\equiv \Tr[\hat \rho_0 \hat f^\dag_\q \hat f_\q]=\frac1{e^{\beta(\eq-\mu)}+1}.
\end{align}
The chemical potential $\mu$ is related to the medium density $n$ via
\begin{align}
    n=\sum_\q n_F(\eq)=\begin{cases}
    -\left(\frac{m}{2\pi \beta}\right)^{3/2}{\rm Li}_{3/2}(-e^{\beta\mu}) & \rm(3D)\\[0.5em] \frac{m}{2 \pi \beta} \ln(1+ e^{\beta \mu}) & 
     \mbox{(2D)}\end{cases}
\end{align}
where ${\rm Li}$ is the polylogarithm. The density is related to the Fermi energy via
\begin{align}
    E_F=\frac{k_F^2}{2m}=\begin{cases}
    \frac{(6\pi^2n)^{2/3}}{2m} & \rm(3D)\\ \frac{4\pi n}{2m} & \rm{(2D)}\end{cases}
\end{align}

\subsection{Impurity spectral function}

In the limit of a weak Rabi coupling, we can obtain the spin-$\up$ impurity spectral function within linear response:
\begin{align}
    A_\up(\omega) \simeq \sum_l |\alpha^{\uparrow (l)}_0|^2 \delta(\omega - E_l),
    \label{eq:spec}
\end{align}
where the variational equations in Eq.~\eqref{eq:VariationalEqs} are solved at $\Omega_0=0$.
In practice, the spin-$\down$ impurity interacts weakly with the medium, and thus the spectral function can be measured by driving transitions from the initial (nearly) non-interacting spin-$\down$ impurity state into the spin-$\up$ state. Indeed, this has been done in both experiments~\cite{Scazza2017,Oppong2019}.

Since the solution of Eq.~\eqref{eq:VariationalEqs} are discrete, we convolve the resulting spectrum with a Gaussian to yield
\begin{align}
    I(\omega) &= \sum_l |\alpha^{\uparrow (l)}_0|^2 g(\omega - E_l).
\end{align}
Convolution of the spectrum in this case has the added benefit of enabling one to approximately model the finite duration of the pulses used in experiment.

\subsection{Simulating Rabi oscillations}

The Rabi oscillations are defined by
\begin{align}
{\cal N}_\down(t)=\expval{\hat c(t)\hat n_\down \hat c^\dag(t)}_\beta.
\end{align}
We will take as our initial condition that $\hat c(\fix{t=0})=\hat c_{\0\down}$, i.e., the impurity is initially in a bare spin-$\downarrow$ state, which means that the impurity number is ${\cal N}_\down+{\cal N}_\up=1$ at all times. The Rabi oscillations are then given by
\begin{align}
{\cal N}_\down(t)& %
\simeq \Trace[\hat\rho_0\hat{c}_\down(t)\hat n_\down\hat{c}^\dag_\down(t)]
= \sum_{j} \fix{\expval*{\hat{O}_j \hat{n}_\down \hat{O}^\dag_j}_\beta} \bigg|\sum_l \alpha_{0}^{\downarrow(l)^*} e^{-i E_l t} \alpha_j^{\downarrow(l)} \bigg|^2  .\label{eq:RabiOscEq}
\end{align}

\begin{figure}
    \centering
    \includegraphics[width=0.4\linewidth]{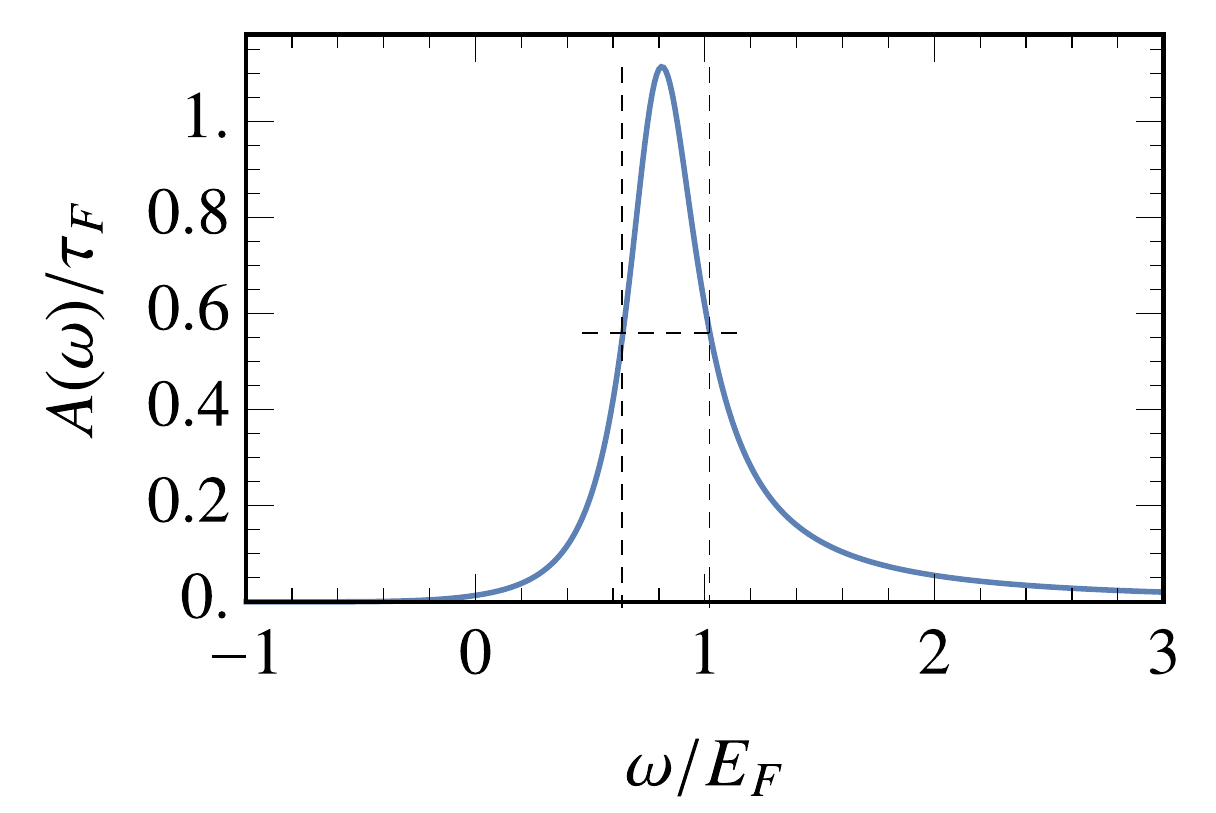}
    \caption{Example of the calculation of the uncertainty in the detuning of Rabi oscillations onto the repulsive polaron. We show the spectral function of the repulsive polaron in the 2D ${}^{173}$Yb experiment of Ref.~\cite{Oppong2019} (solid line), with the dashed lines indicating the full-width half-maxima. The detuning is set %
    between these half-maxima, leading to the uncertainty shown in Fig.~\ref{fig:Rabi} in the main text. The shown spectral function is for $\ln(1/k_F a_{\text{2D}}) = 0.41$, $k_F R_{\text{2D}} = 0.69$ and $T/T_F = 0.16$.}
    \label{fig:DetuningUncertaintyCalc}
\end{figure}

With the exception of detuning, all of the parameters used to define the Rabi oscillations are provided from the relevant experiment. The detuning in experiment is set to address the repulsive polaron peak, and to match this procedure, we use a calculated detuning such that the Rabi oscillations address the theoretically obtained spin-$\up$ repulsive polaron. Due to small but finite initial state interactions, this detuning must take into account the energy of the spin-$\downarrow$ impurities. This is achieved by assuming the spin-$\downarrow$ impurities exist as zero-momentum repulsive polarons with a narrow spectral width. Owing to this assumption, combined with thermal fluctuations, a finite density of impurities and experimental limitations, we place an uncertainty on the detuning (see Fig.~\ref{fig:DetuningUncertaintyCalc}). This requires that we simulate Rabi oscillations over the range of possible detunings.
The parameters used in the simulations of the Rabi oscillations in Figs.~\ref{fig:Rabi} and \ref{fig:ResidueAndDamping} can be found in Table~\ref{tab:params}.

We find that the inclusion of initial state interactions leads to a small reduction in the damping of the Rabi oscillations.
This can be understood from the fact that in the limit in which the spin states have equal interactions, spin-symmetry implies that the Rabi oscillations would be undamped and oscillate at the bare Rabi frequency.

\begin{figure}[th]
    \centering
    \includegraphics[width=\linewidth]{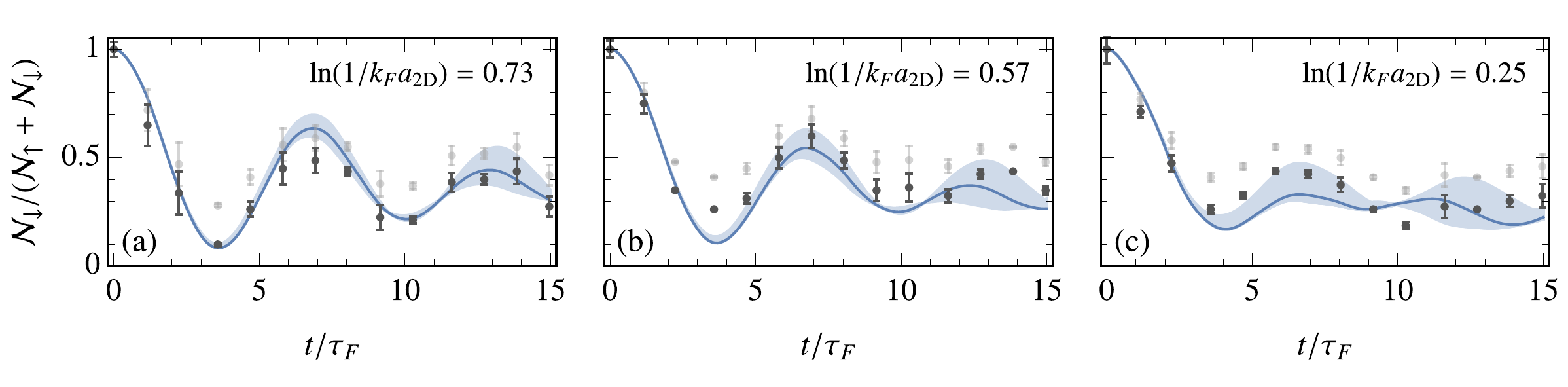}
    \caption{\jfl{Comparison of theory (blue solid lines) of Fig.~2(a)-(c) in the manuscript and the experimental data (black circles) with a constant offset of $0.2$ removed and normalized to the value at time $t=0$.
    The light-gray circles correspond to the unmodified data in the main text.}}
    \label{fig:RabiSamples2D_finiteoffset}
\end{figure}

\jfl{
\subsection{Finite offset in the Rabi oscillation measurement of the 2D experiment}
As noted in the main text, there is a slight disagreement in the amplitude of the Rabi oscillations in the 2D experiment \cite{Oppong2019} and our variational approach.
This disagreement can be explained with a constant offset $\epsilon > 0$ of the $\mathcal{N}_\downarrow$ measurement in the experiment.
In our work, the relative population $\mathcal{N}_\downarrow/(\mathcal{N}_\downarrow + \mathcal{N}_\uparrow)$ of the 2D experiment is inferred from the sole measurement of $\mathcal{N}_\downarrow(t)$ and
\begin{align}
\mathcal{N}_\downarrow\left/(\mathcal{N}_\downarrow + \mathcal{N}_\uparrow)\right. \approx \mathcal{N}_\downarrow\left/[\mathcal{N}_\downarrow(t = 0) + \mathcal{N}_\uparrow(t = 0)]\right. \approx \mathcal{N}_\downarrow/[\mathcal{N}_\downarrow(t = 0) + 0] = \mathcal{N}_\downarrow\left/\mathcal{N}_\downarrow(t = 0)\right..
\end{align}
%
Thus, a constant offset in the measurement of $\mathcal{N}_\downarrow$  changes the relative population as
\begin{align}
\mathcal{N}_\downarrow\left/\mathcal{N}_\downarrow(t = 0)\right. \rightarrow (\mathcal{N}_\downarrow + \epsilon)\left/[\mathcal{N}_\downarrow(t = 0) + \epsilon]\right. = \mathcal{N}_\downarrow\left/[\mathcal{N}_\downarrow(t = 0) + \epsilon]\right. + \epsilon\left/[\mathcal{N}_\downarrow(t = 0) + \epsilon]\right.,
\end{align}
which directly shows how the mean of the relative population is artificially increased by $\epsilon/[\mathcal{N}_\downarrow(t = 0) + \epsilon]$ and the amplitude is artificially reduced by the additional term $\epsilon$ in the denominator.

The offset $\epsilon$ in the experimental data originates from the detection method, for which the majority Fermi sea is removed, but a finite number of remaining majority atoms contributes a positive spurious signal to the measurement of $\mathcal{N}_\downarrow$.
This contribution is independent of $t$ and we estimate $\epsilon\lesssim 0.2\, \mathcal{N}_\downarrow$ from a reevaluation of the existing data of Ref.~\cite{Oppong2019}.
In Fig.~\ref{fig:RabiSamples2D_finiteoffset}, we illustrate how the removal of such an offset considerably 
improves the agreement between experiment and the variational approach.
However, we have chosen not to subtract $\epsilon$ from the data in Fig.~2 of the main text as we lack a precise number for each data set.}

\heavyrulewidth=.08em
\lightrulewidth=.05em
\cmidrulewidth=.03em
\belowrulesep=.65ex
\belowbottomsep=0pt
\aboverulesep=.4ex
\abovetopsep=0pt
\cmidrulesep=\doublerulesep
\cmidrulekern=.5em
\defaultaddspace=.5em

\begin{table}[h]
    \caption{\label{tab:params}%
	    Parameters used for obtaining the theoretical curves and simulations shown in Figs.~2~and~3, matching the experimental ones from Refs.~\cite{Oppong2019} and \cite{Scazza2017}. Experimental uncertainties are given in parenthesis were applicable. Note that the 3D effective range is always zero.\bigskip}
    \begin{tabular}{l l c c c c c c c c c c c} 
        \midrule\midrule
        && \multicolumn{7}{c}{Fig.~2} & \hspace{1.75em} & \multicolumn{3}{c}{Fig.~3} \\ \cmidrule{3-9} \cmidrule{11-13}
        && (a) & (b) & (c) & \hspace{0.5em} & (d) & (e) & (f) & & (a), (c) & \hspace{0.5em} & (b), (d) \\ \midrule 
        Interaction parameter & $\ln(1/k_F a_{\mathrm{2D}\up})$ \hspace{0.75em} & 0.73(4) & 0.57(5) & 0.25(5) & & \multicolumn{3}{c}{---} & & 0.07--0.91 & &  --- \\
        & $\ln(1/k_F a_{\mathrm{2D}\downarrow})$ & \multicolumn{3}{c}{4.9(1)} & & \multicolumn{3}{c}{---} & & 4.9(1) & &  --- \\
        & $1/k_F a_{\mathrm{3D}\up}$ & \multicolumn{3}{c}{---} & & 2.63(4) & 1.27(2) & 0.22(1) & & --- & & 0.22--4.23 \\
        & $1/k_F a_{\mathrm{3D}\downarrow}$ & \multicolumn{3}{c}{---} & & 9.20(15) & 6.94(11) & 5.03(8) & & --- & & 5.03--11.43 \\
        Range parameter & $k_F R_{\text{2D}\up}$ \hspace{1.5em} & 0.71 & 0.67 & 0.68 & & \multicolumn{3}{c}{---} & & 0.67--0.76 & &  --- \\
        Rabi coupling & $\Omega_0/E_F$ & 0.95(11) & 0.96(11) & 0.94(11) && 0.68(1) & 0.69(1) & 0.67(1) & & 1.08 && 0.70 \\
        Reduced temperature & $T/T_F$ & \multicolumn{3}{c}{0.16(4)} && \multicolumn{3}{c}{0.13(2)} & & 0.16(4) && 0.13--0.14 \\
        \addlinespace[0.5em]
        Rep. polaron energy & $E_{+ \uparrow}/E_F$ & 0.71 & 0.76 & 0.89 && 0.19 & 0.42 & 1.08 & & 0.64--1.02 && 0.11--1.08 \\
         & $E_{+ \downarrow}/E_F$ & \multicolumn{3}{c}{0.19} && 0.05 & 0.07 & 0.09 & & 0.19 && 0.04--0.09 \\
        Detuning uncertainty & $\delta E_+/E_F$ & 0.24 & 0.31 & 0.47 && 0.02 & 0.05 & 0.42 & & 0.19--0.65 && 0.02--0.42 \\
        \midrule\midrule
    \end{tabular}
\end{table}

\section{Green's function approach for quasiparticle properties}
A key alternative to our study of the impurity dynamics and properties in the TBM is provided through a Green's function approach. It has been shown that a variational approach using a single particle-hole excitation is equivalent to a Green's function approach calculated with non-self consistent $T$ matrix theory --- see Ref.~\cite{Combescot2007} for a zero-temperature treatment, or Ref.~\cite{Liu2019} at finite temperature. %

In this section, we take $\Omega_0=0$ in the variational equations~\eqref{eq:VariationalEqs}, while we consider the more general case in the next section. This allows us to derive a finite temperature impurity self energy $\Sigma_\sigma(E)$ separately within each of the impurity subspaces. Solving these equations for the energy then yields the expression
\begin{align} \label{eq:definingSelfE}
    E =  \sum_{\q} n_F(\eq) \left[ \frac{E - \varepsilon_{\q \sigma}}{g_\sigma^2} - \sum_{\k} \frac{1 - n_F(\ek)}{E - \varepsilon_{\k \q}} \right]^{-1}.
\end{align}
The right hand side of this expression is precisely the impurity self energy at zero momentum using ladder diagrams at finite temperature~\cite{Liu2019}. The self energy is then related to the impurity (single-particle) Green's function through Dyson's equation,
\begin{align} \label{dyson}
    G_\sigma(E) = \frac{1}{E - \Sigma_\sigma(E)}.
\end{align}

The relevant properties of the repulsive polaron can now be defined in terms of the impurity self energy. In particular, the repulsive polaron energy $E_{+\sigma}$ is a (positive) solution to the implicit equation
\begin{align}
    \Re \left[\Sigma_\sigma (E) \right] = E.
\end{align}
Expanding the Green's function around this pole, the repulsive polaron quasiparticle residue is
\begin{align}
    Z_\sigma = \left( 1 - \left. \pdv{\Re\left(\Sigma_\sigma(E) \right)}{E} \right|_{E = E_{+\sigma}} \right)^{-1},
\end{align}
and the quasiparticle width is
\begin{align} \label{Eq:quasiparticlewidth}
    \Gamma_\sigma = - Z_\sigma \Im \left[ \Sigma_\sigma(E_{+\sigma}) \right].
\end{align}
In addition to these properties, the impurity spectral function in Eq.~\eqref{eq:spec} is given by
\begin{align}
    A_\sigma(E) = -\frac{1}{\pi} \Im[G_\sigma(E)].
\end{align}

\subsection{Repulsive polaron width at weak interactions}

Here we provide details of the calculation of the approximate width of the repulsive polaron peak at weak interactions, Eq.~\eqref{eq:GammaPT}. To perform this calculation, we specialize to three dimensions, zero temperature, and to a broad Feshbach resonance, i.e., $R_{\rm 3D}=0$. In fact, the arguments in the following are valid as long as the thermal wavelength exceeds the scattering length while $R_{\rm 3D}\lesssim 1/(na_{\rm 3D}^2)$. 

Assuming zero impurity momentum, the spin-$\up$ impurity self-energy in Eq.~\eqref{eq:definingSelfE} is given by (for simplicity, in this section we suppress all spin indices as well as ``3D'' subscripts)
\begin{align}
    \Sigma(E) = \sum_{\vectorbold{q}} \Theta(k_F - q) \left[ \frac{m}{4 \pi a} - \left( \sum_{\k} \frac{1}{2 \epsilon_{\k}} + \sum_{\k} \frac{1-\Theta(k_F - k)}{E - \varepsilon_{\k \vectorbold{q}}+i0} \right) \right]^{-1}.
\end{align}
where $k\equiv |\k|$ and $q \equiv |\vectorbold{q}|$ and we take the limit of $R_{\text{3D}}\to 0$, which according to Eq.~\eqref{eq:renorm3D} is equivalent to taking the limit of $\nu,g\to\infty$ in such a way that
\begin{align}
    \frac{\nu}{g^2} %
    =   - \frac{m }{4 \pi a} + \sum_{\vb{k}}^\Lambda \frac{1}{2\epsilon_{\vb{k}} }.
\end{align}
We have also introduced a convergence factor $+i0$ which shifts the energy poles by an infinitesimal amount into the lower half of the complex plane.

In order to find the approximate form of the self-energy (and thereby the width) in the limit of weak interactions, we perturbatively expand the self-energy in scattering length (up to order $a^2$):
\begin{align}
    \Sigma(E) = \sum_{\vectorbold{q}} \Theta(k_F - q) \left[ \frac{4 \pi a}{m} + \frac{16 \pi^2 a^2}{m^2} \left( \sum_{\k} \frac{1}{2 \epsilon_{\k}} + \sum_{\k} \frac{1-\Theta(k_F - k)}{E - \varepsilon_{\k \vectorbold{q}} + i 0} \right) \right].
\end{align}
In the limit of weak repulsive interactions, the repulsive polaron will have residue $Z \simeq 1$ and the width is thus given by
\begin{align}
    \Gamma \simeq - \Im[\Sigma(E_+)].
\end{align}
We can extract the imaginary component of the self-energy through the symbolic identity, which is valid for all real $\alpha$:
\begin{align}
    \frac{1}{\alpha + i 0} = \mathscr{P} \frac{1}{\alpha} - i\pi \delta(\alpha),
\end{align}
where $\mathscr{P}$ denotes the Cauchy principal value. We thus have,
\begin{align}
    - \Im[\Sigma(E)] = \frac{16 \pi^3 a^2}{m^2 E_F} \sum_{\vectorbold{q}, \k } \Theta(k_F - q) (1 - \Theta(k_F - k)) \delta(E/E_F - \varepsilon_{ \k \vectorbold{q}}/E_F).
\end{align}
In the thermodynamic limit, these sums reduce to integrals that can be solved analytically in spherical coordinates:

\begin{align}
    - \Im[\Sigma(E)] = & \frac{k_F^4 a^2}{16 \pi m E_F^2} \Bigg[ 4E^2  \ln \left(\frac{2E_F}{\sqrt{E_F (2   E+E_F)}+E_F}\right) %
    +3E^2-4 E E_F-2
   E_F^2+2 (E+E_F)
   \sqrt{E_F (2 E+E_F)}
    \Bigg].
\end{align}

\begin{figure}
    \centering
    \includegraphics[width=0.4\linewidth]{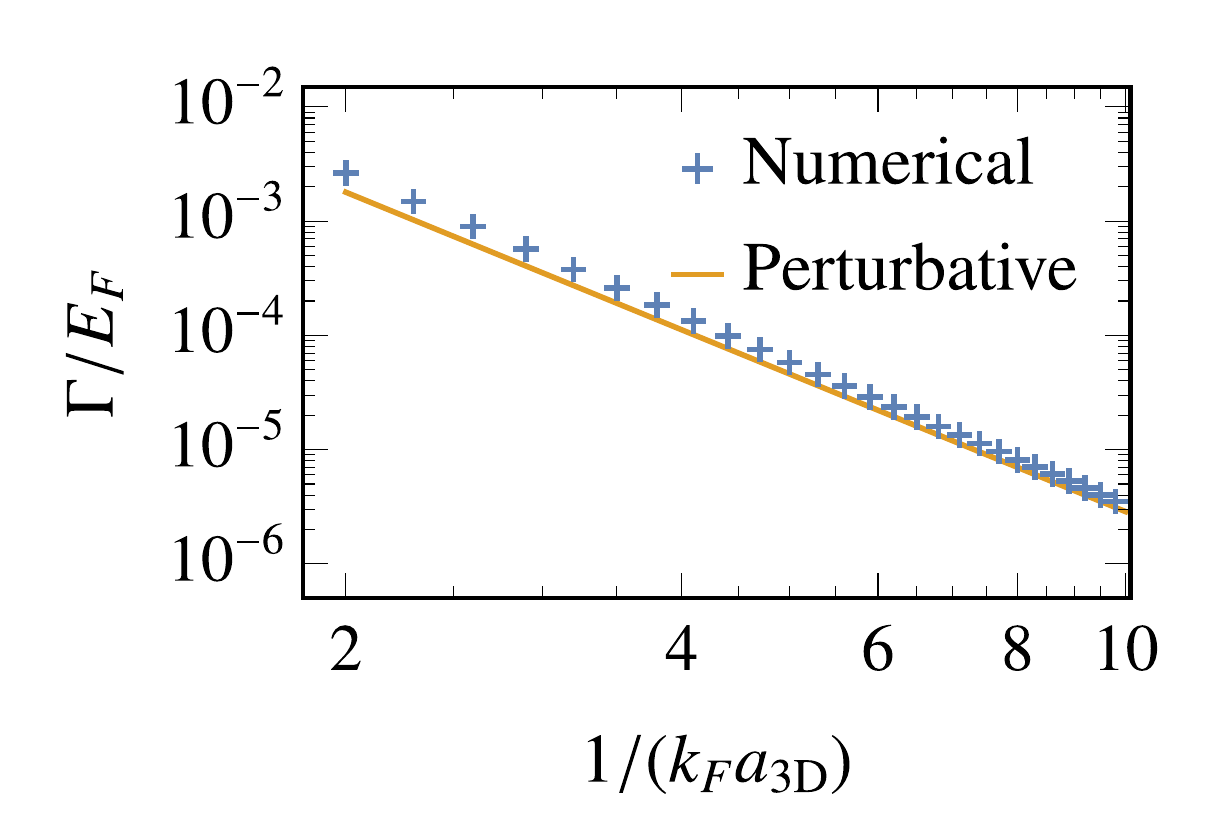}
    \caption{Comparison of the perturbative expression for the repulsive polaron quasiparticle width in Eq.~\eqref{PertbExpDamp} against the numerically exact calculation from Eq.~\eqref{Eq:quasiparticlewidth}.} %
    \label{fig:PertAndNumerical}
\end{figure}

Importantly, the imaginary part of the self energy is zero at zero energy, and we must therefore consider finite energy.
In the limit of weak interactions, the energy of the repulsive polaron is given by the mean-field approximation~\cite{Bishop1973}:
\begin{align}
    E_+ = \frac{2 k_F^3 a}{3 \pi m}.
\end{align}
Using this energy and only retaining terms of order $a^4$ (the lowest non-zero contribution) the width is given by
\begin{align} \label{PertbExpDamp}
    \frac{\Gamma}{E_F}= \frac{8  }{9 \pi^3} (k_F a)^4.
\end{align}
Since we originally expanded the self-energy up to order $a^2$ and have ended with a result that is of order $a^4$ we justify this result numerically in Fig.~\ref{fig:PertAndNumerical}. Furthermore, we have checked that all contributions with multiple particle-hole excitations vanish at order $a^4$.

\section{Approximate Rabi oscillations based on Green's functions}

We finally turn to the arguments that led to Eq.~\eqref{eq:TheoryFit} in the main text, which allowed us to link the repulsive polaron width to the damping of the Rabi oscillations. This will allow us to extend the standard approximation for extracting polaron properties from impurity Rabi oscillations \cite{Kohstall2012}. 
It is useful to introduce a spectral decomposition of the Rabi oscillations, which is provided by the Fourier transform of Eq.~\eqref{eq:RabiOscEq}: %
\begin{align} \label{eq:RabiSpectrum}
    \mathcal{R}_\downarrow (\omega)\equiv \int dt\,e^{i\omega t}{\cal N}_\down(t) \simeq \sum_{j, l,l'} \fix{\expval*{\hat{O}_j \hat{n}_\down \hat{O}^\dag_j}_\beta}  \, \alpha^{\downarrow (l)^*}_{0}  \alpha^{\downarrow(l)}_{j}\alpha^{ \downarrow(l')}_{0} \alpha^{\downarrow(l')^*}_{j} \delta(\omega - E_l + E_{l'}).
\end{align}
If the interactions are not too strong, the sum over intermediate states is dominated by the ``bare'' $\alpha_0$ component. \fix{In this case, $\expval*{\hat{O}_j \hat{n}_\down \hat{O}^\dag_j}_\beta \simeq \expval*{\hat{O}_j \hat{c}^\dag_{\0 \down} \hat{c}_{\0 \down} \hat{O}^\dag_j}_\beta$} and thus
\begin{align} %
    \mathcal{R}_\downarrow (\omega) \simeq \sum_{l,l'} \abs*{\alpha^{\downarrow (l)}_{0}}^2 \abs*{\alpha^{ \downarrow(l')}_{0}}^2 %
    \delta(\omega - E_l + E_{l'}).
\end{align}
We thus recognize the Rabi spectrum as developing according to a convolution of spectral functions \textit{in the presence of} Rabi coupling:
\begin{align}
{\cal N}_\down(t)& \simeq 
\int d \omega\, d \omega' \, \Tilde{A}_\downarrow (\omega) \Tilde{A}_\downarrow (\omega') e^{-i (\omega - \omega')t}.
\end{align}
Here 
\begin{align}
    \Tilde{A}_\downarrow (\omega) = -\frac{1}{\pi} \Im[\Tilde{G}_\downarrow (\omega)],
\end{align}
is calculated from the impurity Green's function including Rabi coupling.

We \fix{can approximate} the Rabi coupled Green's function $\tilde G$ via the relation 
\begin{align} \label{eq:ApproxGreenRelation}
    \Tilde{G}(\omega) \fix{\simeq} \mqty( G_\uparrow^{-1}(\omega) & \Omega_0/2 \\ \Omega_0/2 & G_\downarrow^{-1}(\omega))^{-1}\fix{,}
\end{align}
\fix{where, for ease of notation,}
we define $\Tilde{G}_\downarrow (\omega) \equiv \Tilde{G}_{22}(\omega)$. \fix{We point out that in using Eq.~\eqref{eq:ApproxGreenRelation} to calculate $\Tilde{G}_\downarrow (\omega)$, we are ignoring the coexistence of the spin-$\downarrow$ impurity with excitations of the Fermi gas.}
Approximating the decoupled Green's functions as
\begin{align} 
    G_\sigma (\omega) \simeq \frac{Z_\sigma}{\omega - E_{+\sigma} - \delta_{\sigma \downarrow} \Delta \omega + i \Gamma_\sigma},
\end{align}
we find that
\begin{align} \label{eq:ApproxRabi}
    {\cal N}_\down(t) &\simeq Z_\downarrow^2 e^{- \left(\Gamma _{\downarrow}+\Gamma _{\uparrow}\right) t} \left[\frac{\Gamma
   _{\uparrow}-\Gamma _{\downarrow} }{\sqrt{\Omega _0^2 Z_{\downarrow}
   Z_{\uparrow}-\left(\Gamma _{\uparrow}-\Gamma
   _{\downarrow}\right)^2}} \sin \left(t \sqrt{\Omega _0^2 Z_{\downarrow}
   Z_{\uparrow}-\left(\Gamma _{\uparrow}-\Gamma
   _{\downarrow}\right)^2}\right) \right. \nn\\
   &{}\hspace{2em}+ \left. \left(1-\frac{\Omega _0^2 Z_{\downarrow} Z_{\uparrow}}{2
   \Omega _0^2 Z_{\downarrow} Z_{\uparrow}-2 \left(\Gamma _{\uparrow}-\Gamma
   _{\downarrow}\right)^2}\right) \cos \left(t \sqrt{\Omega _0^2 Z_{\downarrow}
   Z_{\uparrow}-\left(\Gamma _{\uparrow}-\Gamma
   _{\downarrow}\right)^2}\right)+\frac{\Omega _0^2 Z_{\downarrow} Z_{\uparrow}}{2
   \Omega _0^2 Z_{\downarrow} Z_{\uparrow}-2 \left(\Gamma _{\uparrow}-\Gamma
   _{\downarrow}\right)^2}\right].
\end{align}
Here, we have taken $ \Delta \omega = E_{+\up} -E_{+\down}$ 
for simplicity (i.e., on resonance Rabi oscillations). In the cases of interest where $Z_\downarrow$ is slightly below 1, the Rabi oscillations are not normalised. However, this is simply an artefact of our approximation of $G_\downarrow(\omega)$ and is overcome by dividing Eq.~\eqref{eq:ApproxRabi} by $Z_\downarrow^2$.

Equation~\eqref{eq:ApproxRabi} allows us to immediately identify the Rabi frequency and damping
\begin{align}
    \Omega &\simeq \sqrt{\Omega _0^2 Z_{\downarrow} Z_{\uparrow}- \left(\Gamma _{\uparrow}-\Gamma_\downarrow\right)^2},
   \label{eq:OmegaRabi}\\
     \Gamma_R & \simeq \Gamma_\downarrow + \Gamma_\uparrow.
\end{align}
These reduce to $\Omega\simeq \sqrt{\Omega _0^2 Z_{\uparrow}-\Gamma _{\uparrow}^2}$ and $\Gamma_R\simeq \Gamma_\up$ in the case of weak initial state interactions. Equation~\eqref{eq:OmegaRabi} illustrates why a strong Rabi coupling is necessary in order to drive coherent oscillations once $\Gamma_\up$ becomes appreciable, which is why the oscillations are strongly suppressed for the strongest repulsive interactions in the 2D case~(see Fig.~\ref{fig:Rabi}). It also implies that we can estimate the quasiparticle residue by
\begin{align}
    Z_\uparrow \simeq \frac{ \Omega ^2 + \Gamma_\uparrow^2}{\Omega _0^2}.
\end{align}
Finally, assuming weak initial state interactions such that $Z_\down=1$ and $\Gamma_\down=0$, and taking $\Omega_0 Z_\up\gtrsim\Gamma_\up$, we arrive at the form in Eq.~\eqref{eq:MainApproxRabi} of the main text:
\begin{align} 
    {\cal N}_\down(t) & \simeq e^{- \Gamma_\up t} 
    \left[\frac12+\frac12 \cos \left(t \sqrt{\Omega_0^2
   Z_\up-\Gamma_\up^2}\right)\right].
\end{align}

The limiting feature of this effective model comes from our approximation of $\hat n_\down \simeq \hat c_{\0 \downarrow}^\dag \hat c_{\0 \downarrow}$. For any $\Gamma_\uparrow, \Gamma_\downarrow >0$, this approximation will always lead to $\mathcal{N}_\downarrow(t) \to 0$ for large $t$, which does not match the behaviour of Rabi oscillations in experiment or the TBM. However, at the intermediate times considered in Fig.~\ref{fig:Rabi} it provides a good model of the actual oscillations.

\end{document}